\documentclass[12pt,journal,onecolumn]{IEEEtran}
\usepackage{epsfig,rotating,setspace,latexsym,amsmath,epsf,amssymb,amsfonts,bm,theorem,subfigure,epstopdf,graphicx}
\usepackage{cite,authblk}
\usepackage{algorithmic,algorithm}
\usepackage{setspace}
\usepackage{color}
\usepackage{dsfont}
 
\newtheorem{theorem}{Theorem}
\newtheorem{lemma}{Lemma}

\newenvironment{Proof}[1]{\medskip\par\noindent{\bf Proof:\,}\,#1}{{\mbox{\,$\blacksquare$}\par}}
\newcommand{\mathbbm}[1]{\text{\usefont{U}{bbm}{m}{n}#1}} 

\allowdisplaybreaks
 
\begin{document}
 
\title{Age-Aware Gossiping in Network Topologies
\thanks{This paper was presented in part at Allerton Conference, September 2022 \cite{mitra_allerton22}.}}
 
\author{Purbesh Mitra \qquad Sennur Ulukus\\
        \normalsize Department of Electrical and Computer Engineering\\
        \normalsize University of Maryland, College Park, MD 20742\\
        \normalsize  \emph{pmitra@umd.edu} \qquad \emph{ulukus@umd.edu}}
 
\maketitle

\vspace*{-2cm}

\begin{abstract}
We consider a fully-connected wireless gossip network which consists of a source and $n$ receiver nodes. The source updates itself with a Poisson process and also sends updates to the nodes as Poisson arrivals. Upon receiving the updates, the nodes update their knowledge about the source. The nodes gossip the data among themselves in the form of Poisson arrivals to disperse their knowledge about the source. The total gossiping rate is bounded by a constraint. The goal of the network is to be as timely as possible with the source. We propose a scheme which we coin \emph{age sense updating multiple access in networks (ASUMAN)}, which is a distributed opportunistic gossiping scheme, where after each time the source updates itself, each node waits for a time proportional to its current age and broadcasts a signal to the other nodes of the network. This allows the nodes in the network which have higher age to remain silent and only the low-age nodes to gossip, thus utilizing a significant portion of the constrained total gossip rate. We calculate the average age for a typical node in such a network with symmetric settings, and show that the theoretical upper bound on the age scales as $O(1)$. ASUMAN, with an average age of $O(1)$, offers significant gains compared to a system where the nodes just gossip blindly with a fixed update rate, in which case the age scales as $O(\log n)$. Further, we show that this $O(1)$ age performance is sustained if a network has only a fraction of fully-connected edges. However, if the nodes have finite $O(1)$ connectivity, e.g., ring networks, two-dimensional grids, we show that ASUMAN scheme underperforms uniform gossiping, pointing to the need for connectivity with opportunistic gossiping. We improve this performance by introducing a hierarchical structure in the network, which recovers $O(1)$ age scaling under $O(\sqrt{n})$ connected networks. Further, we show how the age of the nodes scale when the cluster heads are finitely connected among themselves, e.g., $O(c)$ age scaling for disconnected and $O(\sqrt{c})$ age scaling for ring-connected cluster heads, where $c$ is the number of clusters. Finally, we show that the $O(1)$ age scaling can be extended to asymmetric settings as well. We give an example of power law arrivals, where nodes' ages scale differently but follow the $O(1)$ bound.
\end{abstract}

\section{Introduction}

Gossiping is a mechanism to disperse information quickly in a network. Each node of the network transmits its own data randomly to its neighboring nodes. This kind of technique is particularly useful in dense distributed sensor networks where a large number of nodes communicate with each other without the presence of a centralized server that schedules transmissions. Although gossiping has been studied extensively \cite{yaron03thesis, shah08monograph, Sanghavi2007GossipFileSplit}, the timeliness of gossiping networks is first analyzed in \cite{yates21gossip}. For measuring the timeliness of a system, the age of information metric has been introduced \cite{kaul11AoI, kosta17AoIbook, Sun2019AgeOI, yatesJSACsurvey}. A disadvantage of the traditional age metric is that it does not take the source's update rate into account; even if the information at the source has not changed, the traditional age metric keeps increasing linearly with time. Thus, optimizing the traditional age metric causes a portion of the resources to be wasted into some transmissions that do not contribute to the timeliness of the system. To circumvent this problem, several extended versions of the traditional age metric have been proposed \cite{yates21gossip, cho3BinaryFreshness, zhong18AoSync, maatouk20AOII, melih2020infocom, Abolhassani21version, wang19counting_process} and used in solving different problems  \cite{melih_cache_TWT, melih20LimitedCache, kaswan2021ISIT,  melih21InfectionTracking}. One such metric is the \textit{version age}, which is introduced as a measure of freshness in \cite{yates21gossip, melih2020infocom, Abolhassani21version}. 

Reference \cite{yates21gossip} uses the version age metric in a gossip network, where the source is updated with rate $\lambda_e$, the source updates a fully-connected network of $n$ nodes with a total update rate of $\lambda$, and each node in the network updates the remaining $n-1$ nodes with a total update rate of $\lambda$. \cite{yates21gossip} shows that the version age of an individual node in such a network scales as $O(\log n)$ with the network size $n$. Some variations of this system model have been studied in \cite{buyukates21CommunityStructure, buyukates22ClusterGossip, kaswan22slicingcoding, kaswan22jamming, kaswan22timestomp, melih2021globecom, mitra_infocom23, kaswan23reliable_source,elmagid_age_distribution,delfani_energy_hervesting}: \cite{buyukates21CommunityStructure, buyukates22ClusterGossip} consider clustered networks with a community structure and show improvements in the version age due to clustering; \cite{kaswan22slicingcoding} considers file slicing and network coding and achieves a version age of $O(1)$ for each node; \cite{kaswan22jamming, kaswan22timestomp} consider version age in the presence of adversarial attacks and investigate how adversarial actions affect the version age; \cite{melih2021globecom} considers the binary freshness metric instead of the version age in gossiping; \cite{mitra_infocom23} considers semi-distributed and fully distributed gossiping schemes for fully-connected networks; \cite{kaswan23reliable_source} investigates how updates from reliable and unreliable sources affect the version age of nodes; \cite{elmagid_age_distribution} uses the moment generating function to characterize the age processes of age-aware gossip networks; and \cite{delfani_energy_hervesting} considers the timeliness of a gossip network with energy harvesting source.

In \cite{yates21gossip}, the gossip rate per node is $\lambda$, and thus, the total gossip rate of the network is $n\lambda$. A downside of the kind of gossiping in \cite{yates21gossip} is that the nodes with staler versions also get to gossip to relatively fresher nodes, which does not actually contribute to the timeliness of the network. Our intuition in this paper is that, if the gossip rate of staler nodes could be assigned (shifted) to fresher nodes instead, then the timeliness of the network could be improved. The challenge is how to implement this intuition in a distributed network where there is no centralized server. 

To that end, we introduce ASUMAN, an age-aware distributed gossiping scheme. Our key idea is reminiscent of the opportunistic channel access scheme proposed in \cite{zhao5SensorNetwork, zhao5opportunistic} in a different context, different system model, with a different goal. \cite{zhao5SensorNetwork, zhao5opportunistic} consider a fading multiple access channel with distributed users. It is well-known \cite{knoppp_humblet_capacity, Tse_Hanly_98_multicaccess} that in a fading multiple access channel, in order to maximize the sum rate, only the largest-channel-gain user should transmit. While the receiver may measure user channel gains and announce the largest-channel-gain user as a feedback in the downlink, the approach in \cite{zhao5SensorNetwork, zhao5opportunistic} is that users measure their own channel gains in the downlink, and apply an opportunistic carrier-sensing-like \cite{CSMA1975} scheme in the uplink. In \cite{zhao5SensorNetwork, zhao5opportunistic}, before starting transmissions, each user waits for a back-off time which is a decreasing function of its own channel gain. Since the user with the largest channel gain waits the least amount of time, it starts transmitting first, all other users become aware of this, and remain silent for the duration of transmission. That is, the broadcast nature of the wireless channel is exploited as an implicit feedback mechanism for the coordination of distributed users.

We use a similar concept in the context of wireless sensor nodes with the objective of information freshness. In our setting, where each node knows its own age, we are interested in enabling the freshest node to capture the channel and update the remaining staler users. In our opportunistic gossiping scheme ASUMAN, each node waits for a back-off time proportional to its own age before starting to gossip. Since the freshest node will start gossiping first, upon hearing this, the rest of the nodes will forgo gossiping for that cycle, and will only potentially receive updates. We show that this policy achieves an age scaling of order of $O(1)$ as the number of nodes increases. For our analysis, we use the stochastic hybrid system (SHS) approach \cite{hespanha_SHS}, similar to \cite{yates21gossip}, to derive the expressions for the average steady-state age values.

Next, we extend this result to a model with partial connectivity, where the nodes are connected to only a fraction of their neighboring nodes. We show that this setting also yields $O(1)$ age performance. Then, we analyze the age performance of a network with finite connectivity and show that this kind of networks perform poorly under ASUMAN. However, we show subsequently that the connectivity order can be reduced to $O(\sqrt{n})$ while keeping $O(1)$ age performance, if we consider hierarchical clusters of fully-connected networks. We study the clustered network topologies in detail for different connectivity levels of the cluster heads, e.g., disconnected and ring-connected cluster heads. Finally, we consider a fully-connected network with asymmetric update rates, with an example of power law arrivals, and derive its average age.

\section{System Model}

\begin{figure}[t]
\centerline{\includegraphics[width=0.5\textwidth]{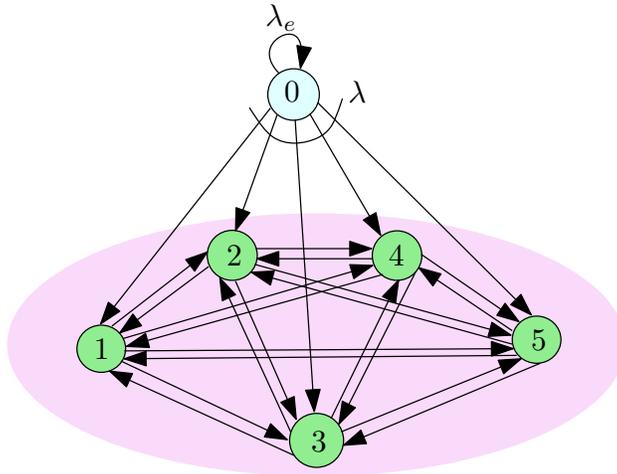}}
\caption{System model for the example case of $n=5$ nodes. Source node $0$ updates itself with rate $\lambda_{e}$ and sends updates to the nodes $\mathcal{N}=\{1,2,3,4,5\}$ uniformly with total rate $\lambda$, i.e., with rate $\lambda/5$ to each of the nodes. The nodes gossip with each other with a total update rate $B=n\lambda=5\lambda$.}
\label{SystemModel}
\vspace*{-0.4cm}
\end{figure}

We consider a system with a source node, labeled as 0, and a set of nodes $\mathcal{N}=\{1,2,\ldots,n\}$; see Fig.~\ref{SystemModel}. The source node updates itself with a Poisson process (i.e., inter-update times are i.i.d.~exponential random variables) with rate $\lambda_{e}$ and it sends updates to each of the nodes in the network as Poisson arrivals with rate $\frac{\lambda}{n}$. The network has a total gossiping rate $B$. The nodes gossip their knowledge about the source's information to maintain the timeliness of the overall network. We consider the version age metric for measuring this timeliness. The version age of the $i$th node, denoted as $\Delta_i(t)$, is the version of information present in the $i$th node as compared to the current version at the source. That is,
\begin{align}
    \Delta_i(t)=N_s(t)-N_i(t),    \label{Version}
\end{align}
where $N_s(t)$ is the version at the source and $N_i(t)$ is the version at the $i$th node at time $t$. We consider the age vector $\mathbf{\Delta}(t)=[\Delta_1(t), \Delta_2(t), \ldots, \Delta_n(t)]$ to denote the version of all the nodes in the network. If the source updates itself at any time, all the elements of $\mathbf{\Delta}(t)$ increase by 1. We assume that the nodes are aware of their own version age. The nodes gossip among themselves to disperse the information in the network. When node $i$ sends a gossip update to node $j$ at time $t$, node $j$ updates its information if the received information is fresher, otherwise it keeps its information as is. Thus, the age of node $j$ is updated to $\Delta'_j(t)=\Delta_{\{i,j\}}(t)=\min \{\Delta_i(t),\Delta_j(t)\}$. 

\section{Opportunistic Gossiping via ASUMAN} \label{sect:asuman}

\begin{figure*}[t]
\centerline{\includegraphics[width=\textwidth]{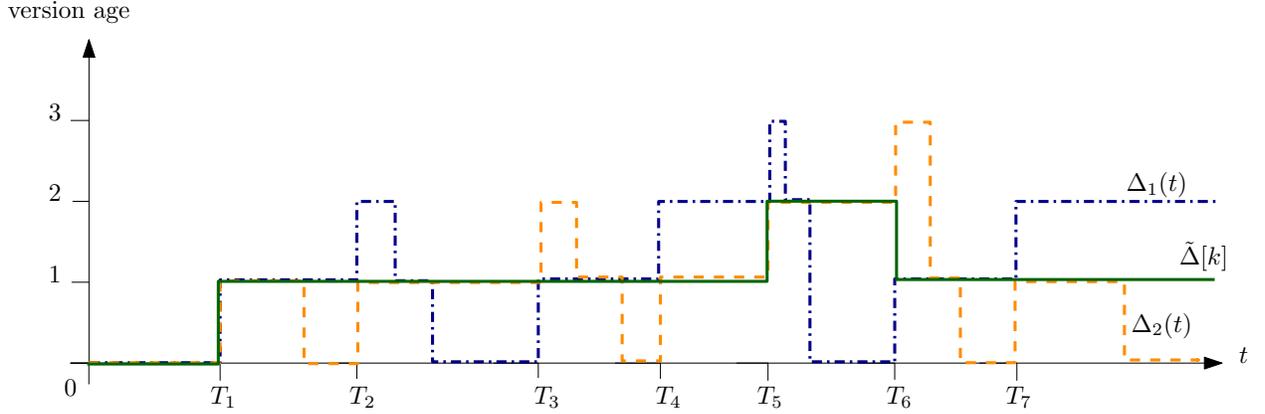}}
\caption{Timeline example for $n=2$. Source $0$ updates itself with rate $\lambda_{e}$ and updates each of the nodes $\mathcal{N}=\{1,2\}$ with rate $\lambda/2$. The version ages of the individual nodes are denoted by $\Delta_1(t)$ and $\Delta_2(t)$, respectively, and $\tilde{\Delta}[k]=\min\{\Delta_1(T_k),\Delta_2(T_k)\}$.}
\label{v_a}
\vspace*{-0.4cm}
\end{figure*}

In this section, we define the ASUMAN scheme and derive a theoretical upper bound for its average age of gossip. Since each node is aware of its own age, when the source updates its information, it acts as a synchronization signal for all the nodes in the network. Suppose we denote the time instances of source self updates as $T_k$, where $k$ is any positive integer. $T_0$ is defined to be $0$. The inter-arrival times $\tau_{k+1}=T_{k+1}-T_k$ are exponentially distributed with mean $\frac{1}{\lambda_{e}}$. When the source updates its information at time $T_k$, each node stops gossiping. The $i$th node waits for a time $C\Delta_i(T_k)$, where $C$ is a small proportionality constant. After waiting this time, node $i$ broadcasts a signal to all the nodes in the network and starts gossiping. However, if a node receives a broadcast from another node before its waiting period expires, then it remains silent for the time interval $\mathcal{I}_k=[T_k,T_{k+1})$. Thus, for each time interval, only the nodes which have the lowest age at the beginning of the interval get to gossip. At time $T_k$, we use $\mathcal{M}_k$ to denote the set of indices of the nodes with the minimum age, $\tilde{\Delta}[k]=\min_{i} \Delta_i(T_k)$; see Fig.~\ref{v_a}. 

From the broadcast signals, all the nodes in the network get to know that there are total $|\mathcal{M}_k|$ number of minimum-age nodes at time $T_k$. Therefore, each of the nodes in $\mathcal{M}_k$ utilizes only $\frac{B}{|\mathcal{M}_k|}$ of total gossip rate, while all the other nodes do not use any update rate for $\mathcal{I}_k$. If $\tau_{k+1}>C\tilde{\Delta}[k]$, each node in $\mathcal{M}_k$ gossips to every other node with rate $\frac{B}{|\mathcal{M}_k|(n-1)}$ for the time interval $[T_k+C\tilde{\Delta}[k],T_{k+1})$. Otherwise, the source updates itself before the nodes get a chance to gossip opportunistically, and the next interval begins with the same scheme. In this work, we are interested in the steady-state mean of the version age of a node, which is defined as
\begin{align}
    a_i=\lim_{t\to\infty}a_i(t)=\lim_{t\to\infty}\mathbb{E}[\Delta_i(t)].\label{avg_notation}
\end{align}

To evaluate this steady-state mean age, we define the mean of $\tilde{\Delta}[k]$ as $\tilde{a}[k]=\mathbb{E}[\tilde{\Delta}[k]]$ and evaluate it in Lemma~\ref{lemma1}.

\begin{lemma}\label{lemma1}
The mean of minimum age in interval $\mathcal{I}_k$ is  
\begin{align}
\tilde{a}[k]=\sum_{\ell=0}^{k-1}\left(\frac{\lambda_{e}}{\lambda_{e}+\lambda}\right)^{\ell}, \quad k\geq 1.
\end{align}
\end{lemma}

\begin{Proof}
We use induction for the proof. Since all the ages are 0 at the beginning, $\tilde{a}[0]=0$ and $\tilde{a}[1]=1$. Assume that the statement is true for $k$. The probability that the source does not update any node in $\mathcal{I}_k$ is $e^{-\lambda\tau_{k+1}}$. If any node in the network is updated in $\mathcal{I}_k$, $\tilde{\Delta}[k+1]$ becomes 1; otherwise it is $\tilde{\Delta}[k]+1$. Thus, we have
\begin{align}
    \tilde{a}[k+1]=\mathbb{E}[(1-e^{-\lambda \tau_{k+1}})+(\tilde{\Delta}[k]+1)e^{-\lambda \tau_{k+1}}].
\end{align}
Since $\tau_{k+1}$ is exponentially distributed with parameter $\lambda_{e}$,
\begin{align}
    \mathbb{E}\left[e^{-\lambda \tau_{k+1}}\right]=\int_0^{\infty}e^{-\lambda \tau_{k+1}}\lambda_{e}e^{-\lambda_{e} \tau_{k+1}}d\tau_{k+1}=\frac{\lambda_{e}}{\lambda_{e}+\lambda}.
\end{align}
Thus, we obtain
\begin{align}
    \tilde{a}[k+1]=1+\tilde{a}[k]\frac{\lambda_{e}}{\lambda_{e}+\lambda}.\label{recurrence}
\end{align}
Now, using the induction hypothesis, we can rewrite \eqref{recurrence} as
\begin{align}
    \tilde{a}[k+1]=1+\sum_{\ell=0}^{k-1}\left(\frac{\lambda_{e}}{\lambda_{e}+\lambda}\right)^{\ell}\cdot\frac{\lambda_{e}}{\lambda_{e}+\lambda}=\sum_{\ell=0}^{k}\left(\frac{\lambda_{e}}{\lambda_{e}+\lambda}\right)^{\ell},
\end{align}
completing the proof.
\end{Proof}

Next, in Lemma~\ref{lemma2}, we consider the idealistic case of $C=0$, i.e., all the nodes instantaneously know about the minimum age nodes in the beginning of the interval $\mathcal{I}_k$. Although this is not a feasible model, the result of this lemma will be used for calculations in the case of $C>0$. In addition, in Lemma~\ref{lemma2}, we consider the case where the total gossip  rate of the network is $B=n \lambda$, which is the same as the total gossip rate in \cite{yates21gossip}.  

\begin{lemma}\label{lemma2}
For $C = 0$, if the total gossip rate is $B=n\lambda$, then the steady-state mean version age of a node scales as $O(1)$.
\end{lemma}

\begin{Proof}
Since the system is symmetric with respect to any node in the network, proving the result only for any fixed $i$th node will suffice. To analyze the system, we follow the SHS formulation in \cite{hespanha_SHS}. Since $C=0$, only one type of state transition is involved. Thus, $\mathcal{Q}=\{0\}$ and for the node $i\in\mathcal{N}$, we choose a function $\psi_i:\mathbbm{R}^n\times[0,\infty)\to\mathbbm{R}$, such that
\begin{align}
    \psi_i(\mathbf{\Delta}(t),t)=\Delta_i(t).   
\end{align}
Following \cite[Thm.~1]{hespanha_SHS}, we write the expected value of the extended generator function as
\begin{align}
    \mathbb{E}[(L\psi_i)(\mathbf{\Delta}(t),t)]=\sum_{(j,\ell)\in \mathcal{L}}\lambda_{j,\ell}(\mathbf{\Delta}(t),t)\mathbb{E}[\psi_i(\phi_{j,\ell}(\mathbf{\Delta}(t),t))-\psi_i(\mathbf{\Delta}(t),t)],
\end{align}
where $\mathcal{L}$ is the set of all possible state transitions. Define reset maps $\phi_{j,\ell}(\mathbf{\Delta}(t),t)=\hat{\mathbf{\Delta}}(t)=[\hat{\Delta}_1(t),\hat{\Delta}_2(t),\ldots,\hat{\Delta}_n(t)]$ as
\begin{align}
    \hat{\Delta}_i(t)=
    \left\{
	\begin{array}{ll}
		\Delta_i(t)+1,  &\mbox{if } j = 0, \ell=0 \\
		0, &\mbox{if } j=0, \ell=i\\
		\min(\Delta_j(t),\Delta_\ell(t)),  & \mbox{if } j\in\mathcal{N},\ell=i \\
		\Delta_i(t), &\mbox{otherwise}.
	\end{array}
\right.
\end{align}
The update rates $\lambda_{j,\ell}$ are given as
\begin{align}
    \lambda_{j,\ell}(\mathbf{\Delta}(t),t)=
    \left\{
	\begin{array}{ll}
		\lambda_{e}, &\mbox{if } j = 0,\ell=0 \\
		\frac{\lambda}{n}, &\mbox{if } j=0,\ell=i\\
		\lambda_{j,\ell}^{(k)}(t), &\mbox{otherwise},
	\end{array}
\right.
\end{align}
where $\lambda_{j,\ell}^{(k)}(t)$ is the gossip rate of node $j$ to node $\ell$ in the time interval $\mathcal{I}_k$. Since $C=0$,
\begin{align}
    \lambda_{j,\ell}^{(k)}(t)=\frac{B}{|\mathcal{M}_k|(n-1)}\mathbbm{1}\{j\in\mathcal{M}_k, \ell\in\mathcal{N}, t\in\mathcal{I}_k\},
\label{GossipRate1}
\end{align} 
where $\mathbbm{1}\{\cdot\}$ is the indicator function. We can rewrite the expected value of the extended generator function as
\begin{align}
    \mathbb{E}[(L\psi_{i})(\mathbf{\Delta}(t),t)]=&\mathbb{E}\bigg[\lambda_{e}(\Delta_i(t)+1-\Delta_i(t))+\frac{\lambda}{n}(0-\Delta_i(t))\notag\\
    &\qquad +\sum_{j\in \mathcal{N}}\lambda_{j,i}(\mathbf{\Delta}(t),t)\left(\Delta_{\{j,i\}}(t)-\Delta_i(t)\right)\bigg].
\end{align}
Now, for $t\in\mathcal{I}_k$, we write the expectation as 
\begin{align}
   \mathbb{E}[(L\psi_{i})(\mathbf{\Delta}(t),t)]=\lambda_{e}-\frac{\lambda}{n}a_i(t)+\mathbb{E}\bigg[\sum_{j\in \mathcal{M}_k}\lambda_{j,i}^{(k)}(t)(\Delta_{\{j,i\}}(t)-\Delta_i(t))\bigg].\label{GossipEquation}
\end{align}

\begin{figure*}[t]
\centerline{\includegraphics[width=0.94\textwidth]{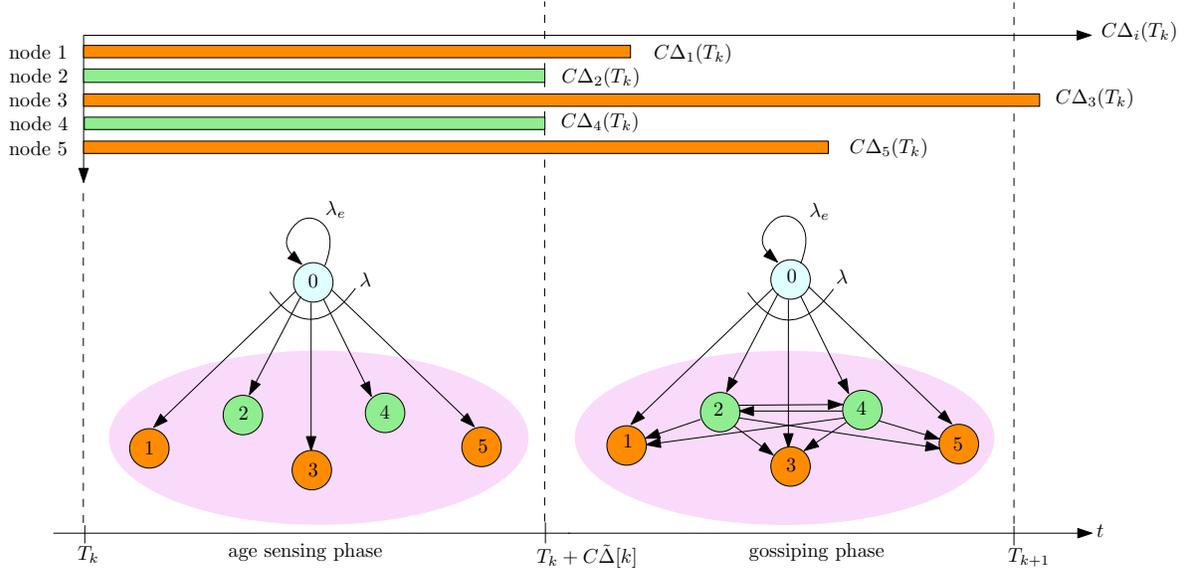}}
\caption{An example of a typical opportunistic gossiping in a $n=5$ node network. At time $T_k$, the minimum age nodes are $\mathcal{M}_k=\{2,4\}$. Thus, they wait for a time of $C\tilde{\Delta}[k]$, where $\tilde{\Delta}[k]=\min_{i} \Delta_i(T_k)$, in the age sensing phase $[T_k, T_k+C\tilde{\Delta}[k])$ and start transmitting with total update rate $B$ in the gossiping phase $[T_k+C\tilde{\Delta}[k],T_{k+1})$.}
\label{illustration}
\vspace*{-0.4cm}
\end{figure*}

Note that, \eqref{GossipRate1} is true, if the $i$th node is not in $\mathcal{M}_k$ by the formulation of our proposed gossiping scheme. However, even if node $i$ is in $\mathcal{M}_k$ we can still assume that it is gossiping to itself with rate $\frac{B}{|\mathcal{M}_k|(n-1)}$, since the corresponding product term $(\Delta_{\{i,i\}}(t)-\Delta_i(t))=0$. Now, since the version age is a piece-wise constant function of time, we obtain
\begin{align}
    \frac{d\mathbb{E}[\psi_i(\mathbf{\Delta}(t),t)]}{dt}=\frac{d\mathbb{E}[\Delta_i(t)]}{dt}=0
\end{align}
for all the continuity points. Hence, the expected value in \eqref{GossipEquation} is $0$, by Dynkin’s formula, as given in \cite{hespanha_SHS}. Thus, \eqref{GossipEquation} becomes
\begin{align}
    0=\lambda_{e}-\frac{\lambda}{n}a_i(t)+\mathbb{E}\bigg[\sum_{j\in \mathcal{M}_k}\frac{B}{|\mathcal{M}_k|(n-1)}\left(\Delta_{\{j,i\}}(t)-\Delta_i(t)\right)\bigg].
\end{align}
Hence, the mean age of an individual node is expressed as
\begin{align}
    \left(\frac{\lambda}{n}+\frac{B}{n-1}\right)a_i(t)=\lambda_e+\mathbb{E}\bigg[\sum_{j\in \mathcal{M}_k}\frac{B}{|\mathcal{M}_k|(n-1)}\Delta_{\{j,i\}}(t)\bigg].\label{j_dep}
\end{align}
In \eqref{j_dep}, $\mathcal{M}_k$ is a function of $\mathbf{\Delta}(t)$. Instead of deriving the distribution of $\mathbf{\Delta}(t)$, we use the inequality $\Delta_{\{j,i\}}(t)\leq \tilde{\Delta}[k]$ for $t\in\mathcal{I}_k$, and rewrite \eqref{j_dep} as the following upper bound 
\begin{align}
    a_i(t)&\leq\frac{\lambda_{e}+\frac{B}{n-1}\mathbb{E}\left[\sum_{j\in \mathcal{M}_k}\frac{\tilde{\Delta}[k]}{|\mathcal{M}_k|}\right]}{\frac{\lambda}{n}+\frac{B}{n-1}},\quad\forall t \in \mathcal{I}_k\\
    &=\frac{\lambda_{e}+\frac{B}{n-1}\tilde{a}[k]}{\frac{\lambda}{n}+\frac{B}{n-1}},\quad\forall t \in \mathcal{I}_k.\label{UB1}
\end{align}
We are interested in the steady-state average age, i.e., average age at $t\to\infty$. We evaluate the asymptote of the upper bound in \eqref{UB1} as $k\to\infty$. From Lemma~\ref{lemma1}, we have 
\begin{align}
    \lim_{k\to \infty}\tilde{a}[k]=\frac{\lambda_{e}+\lambda}{\lambda}.\label{steady_state}
\end{align}

Using \eqref{steady_state} in \eqref{UB1}, we obtain
\begin{align}
    a_i=\lim_{t\to\infty}a_i(t)\leq\lim_{k\to\infty}\frac{\lambda_{e}+\frac{B}{n-1}\tilde{a}[k]}{\frac{\lambda}{n}+\frac{B}{n-1}}=\frac{\lambda_{e}+\frac{B}{n-1} \frac{\lambda_{e}+\lambda}{\lambda}}{\frac{\lambda}{n}+\frac{B}{n-1}}.\label{UB2}
\end{align}
Now, to calculate the scaling of the average age, we use the relation that $B=n\lambda$, which yields
\begin{align}
    \lim_{n\to\infty}a_i\leq\lim_{n\to\infty}\frac{\lambda_{e}}{\lambda}\frac{(1+\frac{n\lambda}{n-1}(\frac{1}{\lambda}+\frac{1}{\lambda_{e}}))}{\left(\frac{1}{n}+\frac{n}{n-1}\right)}=2\frac{\lambda_{e}}{\lambda}+1,
\end{align}
concluding the proof.
\end{Proof} 

Now, we use the results of Lemmas~\ref{lemma1} and \ref{lemma2} to formulate the average version age in Theorem~\ref{thm1}.

\begin{theorem}\label{thm1}
For $C>0$, if $C$ is chosen such that it is bounded for all $n$ and $C\to 0$ as $n\to\infty$, keeping the total gossip rate the same as before, i.e., $B=n\lambda$, then the average version age of a node scales as $O(1)$, and the asymptotic upper bound is the same as that for the case of $C=0$.   
\end{theorem}

\begin{Proof}
There is no change in the function $\psi_i$ or in the reset maps $\phi_{j,\ell}$. The only change is in the update frequencies. For any choice of $C>0$, we divide the time interval $\mathcal{I}_k$ into two phases, an age sensing phase $\mathcal{I}^{(s)}_k=[T_k,\min(T_k+C\tilde{\Delta}[k],T_{k+1}))$ and a gossiping phase $\mathcal{I}^{(g)}_k=[\min(T_k+C\tilde{\Delta}[k],T_{k+1}),T_{k+1})$; see Fig.~\ref{illustration}. 

We already have an upper bound for the $k$th time-slot average age expression for $\mathcal{I}^{(g)}_k$ from Lemma~\ref{lemma2}. Let us denote the right hand side in \eqref{UB1} as $a^{(g)}[k]$, i.e., 
\begin{align}
a^{(g)}[k]=\frac{\lambda_{e}+\frac{B}{n-1}\tilde{a}[k]}{\frac{\lambda}{n}+\frac{B}{n-1}}.
\end{align}

For $\mathcal{I}^{(s)}_k$, we evaluate an upper bound by ignoring the source to $i$th node updates and only considering the opportunistic gossiping. We define the process $\Delta^{(s)}[k]$, such that $\Delta^{(s)}[1]=1$. If the $(k-1)$th interval does not have a gossiping phase, i.e., $\tau_k\leq C\tilde{\Delta}[k-1]$, or if none of the active nodes in $\mathcal{M}_{k-1}$ gossip to node $i$ in $\mathcal{I}^{(g)}_{k-1}$, then $\Delta^{(s)}[k]=\Delta^{(s)}[k-1]+1$. Otherwise, if any node in $\mathcal{M}_{k-1}$ gossips to node $i$ in the interval $\mathcal{I}^{(g)}[k]$, then $\Delta^{(s)}[k]=\tilde{\Delta}[k-1]+1$. We express the probabilistic recurrence relations for $k>1$ as
\begin{align}
    \Delta^{(s)}[k]=
\left\{
	\begin{array}{ll}
		\Delta^{(s)}[k-1]+1, &\mathbb{P}(\tau_{k}\leq C\tilde{\Delta}[k-1])\\
		\Delta^{(s)}[k-1]+1,  &\mathbb{P}(\tau_k > C\tilde{\Delta}[k-1]) e^{-\frac{B}{n-1}(\tau_k-C\tilde{\Delta}[k-1])}\\
		\tilde{\Delta}[k-1]+1, &\mathbb{P}(\tau_k > C\tilde{\Delta}[k-1]) \left(1-e^{-\frac{B}{n-1}(\tau_k-C\tilde{\Delta}[k-1])}\right)
	\end{array}
\right.\label{delta_s}
\end{align}
Clearly, $\Delta_i(t)\leq\Delta^{(s)}[k]$ for $t\in\mathcal{I}^{(s)}_k$. We write the mean of this upper bound as
\begin{align}
    a^{(s)}[k]=& \mathbb{E}\left[\Delta^{(s)}[k]\right]\notag\\
     = &\mathbb{E}\bigg[(\Delta^{(s)}[k-1]+1)\bigg(\mathbb{P}(\tau_k\leq C\tilde{\Delta}[k-1])+\mathbb{P}(\tau_k > C\tilde{\Delta}[k-1])e^{-\frac{B}{n-1}(\tau_k-C\tilde{\Delta}[k-1])}\bigg)\notag\\
    & \quad +(\tilde{\Delta}[k-1]+1)\mathbb{P}(\tau_k > C\tilde{\Delta}[k-1])\left(1-e^{-\frac{B}{n-1}(\tau_k-C\tilde{\Delta}[k-1])}\right)\bigg].\label{SensinUB0}
\end{align}
Since $\tau_k$ is exponentially distributed, we rewrite \eqref{SensinUB0} as
\begin{align}
    a^{(s)}[k]=& (a^{(s)}[k-1]+1)\mathbb{E}\left[1-e^{-\lambda_e C\tilde{\Delta}[k-1]}+e^{-\lambda_e C\tilde{\Delta}[k-1]}\cdot e^{-\frac{B}{n-1}(\tau_k-C\tilde{\Delta}[k-1])}\right]\notag\\
    & +(\tilde{a}[k-1]+1)\mathbb{E}\bigg[e^{-\lambda_e C\tilde{\Delta}[k-1]}\left(1-e^{-\frac{B}{n-1}(\tau_k-C\tilde{\Delta}[k-1])}\right)\bigg].\label{SensingUB}
\end{align}
Now, for $B=n\lambda$, as $n\to \infty$, $C\to 0$, $\frac{B}{n-1}\to \lambda$, $e^{-\lambda_e C\tilde{\Delta}[k-1]}\to 1$, and $e^{\frac{B}{n-1} C\tilde{\Delta}[k-1]}\to 1$. Thus, \eqref{SensingUB} becomes 
\begin{align}
    \lim_{n\to\infty}a^{(s)}[k]=&\left(\lim_{n\to\infty}a^{(s)}[k-1]+1\right)\mathbb{E}\left[e^{-\lambda\tau_k}\right]+\left(\tilde{a}[k-1]+1\right)\mathbb{E}\left[1-e^{-\lambda\tau_k}\right]. \label{recursive_eqn}
\end{align}
From Lemma~\ref{lemma1}, we know that $\mathbb{E}\left[e^{-\lambda\tau_k}\right]=\frac{\lambda_e}{\lambda_e+\lambda}$. Lemma~\ref{lemma1} also says
\begin{align}
    \tilde{a}[k-1]=\sum_{\ell=0}^{k-2}\left(\frac{\lambda_{e}}{\lambda_{e}+\lambda}\right)^{\ell}&\leq \lim_{k\to\infty}\sum_{\ell=0}^{k-1}\left(\frac{\lambda_{e}}{\lambda_{e}+\lambda}\right)^{\ell}=\frac{\lambda_e+\lambda}{\lambda}.\label{a_tilde_UB}
\end{align}
Using \eqref{a_tilde_UB} in \eqref{recursive_eqn} gives
\begin{align}
    \lim_{n\to\infty}a^{(s)}[k]\leq 2+\frac{\lambda_e}{\lambda_e+\lambda}\lim_{n\to\infty}a^{(s)}[k-1]. \label{recursive_eqn2}
\end{align}

Here, we define a new sequence $b[k]$, such that $b[1]=\lim_{n\to\infty}a^{(s)}[1]=1$, and evolves as
\begin{align}
    b[k]=2+\frac{\lambda_e}{\lambda_e+\lambda}b[k-1]. \label{recursive_eqn3}
\end{align}
Therefore, $b[k]\geq \lim_{n\to\infty}a^{(s)}[k]$ for all $k$. Now, using similar logic as in Lemma~\ref{lemma1} here, we obtain an expression for $b[k]$ as 
\begin{align}
    b[k]=2\sum_{\ell=0}^{k-2}\left(\frac{\lambda_{e}}{\lambda_{e}+\lambda}\right)^{\ell}+\left(\frac{\lambda_{e}}{\lambda_{e}+\lambda}\right)^{k-1}.\label{b_k}
\end{align}
Hence, we obtain the relation 
\begin{align}
    \lim_{n\to\infty}a^{(s)}[k]\leq 2\sum_{\ell=0}^{k-2}\left(\frac{\lambda_{e}}{\lambda_{e}+\lambda}\right)^{\ell}+\left(\frac{\lambda_{e}}{\lambda_{e}+\lambda}\right)^{k-1}.\label{a_k}
\end{align}

From \eqref{a_k}, we conclude that $a^{(s)}[k]\sim O(1)$. Note that, since $\Delta_i(t)\leq\Delta^{(s)}[k]$ for $t\in\mathcal{I}_k^{(s)}$, the gossiping process in $\mathcal{I}_k^{(g)}$ cannot increase the age. Thus, $\Delta_i(t)\leq\Delta^{(s)}[k]$ for all $t\in\mathcal{I}_k$. Therefore, $a_{i}(t)\sim O(1)$. This finishes the first part of the statement of Theorem~\ref{thm1}, which is that the asymptotic upper bound for the average age is $O(1)$. 

To prove the next part of the theorem, i.e., that under the conditions given in the statement of the theorem, the upper for $C>0$ is the same as the upper bound for $C=0$, first we note that \eqref{a_k} yields the following steady-state upper bound
\begin{align}
    \!\!\!\lim_{k\to\infty}2\sum_{\ell=0}^{k-2}\left(\frac{\lambda_{e}}{\lambda_{e}+\lambda}\right)^{\ell}+\left(\frac{\lambda_{e}}{\lambda_{e}+\lambda}\right)^{k-1}\!\!=2\left(\frac{\lambda_e}{\lambda}+1\right). \!\!
\end{align}
Comparing to the case of $C=0$, we see that this value is $\lim_{k\to\infty}a^{(g)}[k]+1$. To get a tighter upper bound, we take the age reduction in the gossiping phase into consideration. We have
\begin{align}
    a_i(t)\leq
\left\{
	\begin{array}{ll}
		a^{(s)}[k],  &\text{if } \tau_{k+1}\leq C\tilde{\Delta}[k] \ \forall t \in \mathcal{I}_k\\
		a^{(s)}[k],  &\text{if } \tau_{k+1}> C\tilde{\Delta}[k] \ \forall t \in \mathcal{I}^{(s)}_k\\
		a^{(g)}[k], &\text{if } \tau_{k+1}> C\tilde{\Delta}[k] \ \forall t \in \mathcal{I}^{(g)}_k
	\end{array}
\right.\label{ineq}
\end{align}
We calculate the average age as
\begin{align}
    a_i=\lim_{T\to\infty}\frac{1}{T}\int_{0}^{T}\Delta_i(t)dt=\lim_{T\to\infty}\frac{1}{T}\sum_{k=1}^{N(T)}\beta_i[k], \label{avg_age}
\end{align}
where we denote the number of source self updates as $N(T)=\max\{j:T_j\leq T\}$ and $\beta_i[k]=\int_{\mathcal{I}_k}\Delta_i(t)dt$. Assuming ergodicity of the process, we rewrite \eqref{avg_age} as
\begin{align}
    a_i=\lim_{T\to\infty}\frac{\frac{1}{N(T)}\sum_{k=1}^{N(T)}\beta_i[k]}{T/N(T)}=\frac{\lim_{k\to\infty}\mathbb{E}\left[\beta_i[k]\right]}{\lim_{T\to\infty}T/N(T)}, \label{avg_age2}
\end{align}
if $\lim_{k\to\infty}\mathbb{E}\left[\beta_i[k]\right]$ converges. Since the source self update is a Poisson process with rate $\lambda_e$, $\lim_{T\to\infty}T/N(T)=\frac{1}{\lambda_e}$. We write the numerator of \eqref{avg_age2} as
\begin{align}
    \mathbb{E}\left[\beta_i[k]\right]=&\mathbb{E}\left[\int_{\mathcal{I}_k}\Delta_i(t)\mathbbm{1}\{\tau_{k+1}\leq C\tilde{\Delta}[k]\}dt\right]+\mathbb{E}\left[\int_{\mathcal{I}_k}\Delta_i(t)\mathbbm{1}\{\tau_{k+1}> C\tilde{\Delta}[k]\}dt\right],\label{avg_age3}
\end{align}
The first term in \eqref{avg_age3} constitutes the event when the source updates too quickly for the network to get into the gossiping phase. It is bounded as
\begin{align}
    \mathbb{E}\left[\int_{\mathcal{I}_k}\Delta_i(t)\mathbbm{1}\{\tau_{k+1}\leq C\tilde{\Delta}[k]\}dt\right]
    &=\mathbb{E}_{\tilde{\Delta}[k]}\left[\mathbb{E}\left[\int_{\mathcal{I}_k}\Delta_i(t)\mathbbm{1}\{\tau_{k+1}\leq C\tilde{\Delta}[k]\}dt\bigg|\tilde{\Delta}[k]\right]\right]\\
    &\leq\mathbb{E}_{\tilde{\Delta}[k]}\left[a^{(s)}[k]\mathbb{E}_{\tau_{k+1}}\left[\tau_{k+1}\mathbbm{1}\{\tau_{k+1}\leq C\tilde{\Delta}[k]\}\bigg|\tilde{\Delta}[k]\right]\right].\label{term1}
\end{align}
We obtain the inner expectation in \eqref{term1} as 
\begin{align}
    \mathbb{E}_{\tau_{k+1}}\left[\tau_{k+1}\mathbbm{1}\{\tau_{k+1}\leq C\tilde{\Delta}[k]\}\bigg|\tilde{\Delta}[k]\right]&=\int_{0}^{C\tilde{\Delta}[k]}\tau_{k+1}\lambda_e e^{-\lambda_e\tau_{k+1}}d\tau_{k+1}\\
    &=\frac{1}{\lambda_e}\left(1-e^{-\lambda_e C\tilde{\Delta}[k]}(\lambda_e C\tilde{\Delta}[k] + 1)\right).
\end{align}
For the second term in \eqref{avg_age3}, we break the integral into age sensing and gossiping phases as follows
\begin{align}
    &\mathbb{E}\left[\int_{\mathcal{I}_k}\Delta_i(t)\mathbbm{1}\{\tau_{k+1}> C\tilde{\Delta}[k]\}dt\right]\notag\\
    &=\mathbb{E}_{\tilde{\Delta}[k]}\bigg[\mathbb{E}\bigg[\int_{\mathcal{I}^{(s)}_k}\Delta_i(t)\mathbbm{1}\{\tau_{k+1}> C\tilde{\Delta}[k]\}dt+\int_{\mathcal{I}^{(g)}_k}\Delta_i(t)\mathbbm{1}\{\tau_{k+1}> C\tilde{\Delta}[k]\}dt\bigg|\tilde{\Delta}[k]\bigg]\bigg]\\
    &\leq \mathbb{E}_{\tilde{\Delta}[k]}\bigg[a^{(s)}[k]\mathbb{E}_{\tau_{k+1}}\bigg[C\tilde{\Delta}[k]\mathbbm{1}\{\tau_{k+1}> C\tilde{\Delta}[k]\}\bigg|\tilde{\Delta}[k]\bigg]\bigg] \notag\\
    &\ \ \ +\mathbb{E}_{\tilde{\Delta}[k]}\bigg[a^{(g)}[k]\mathbb{E}_{\tau_{k+1}}\bigg[(\tau_{k+1}-C\tilde{\Delta}[k])\mathbbm{1}\{\tau_{k+1}> C\tilde{\Delta}[k]\}\bigg|\tilde{\Delta}[k]\bigg]\bigg]\\
    &=\mathbb{E}_{\tilde{\Delta}[k]}\bigg[(a^{(s)}[k]-a^{(g)}[k])\mathbb{E}_{\tau_{k+1}}\bigg[C\tilde{\Delta}[k]\mathbbm{1}\{\tau_{k+1}> C\tilde{\Delta}[k]\}\bigg|\tilde{\Delta}[k]\bigg]\bigg]\notag\\
    &\ \ \ +\mathbb{E}_{\tilde{\Delta}[k]}\bigg[a^{(g)}[k]\mathbb{E}_{\tau_{k+1}}\bigg[\tau_{k+1}\mathbbm{1}\{\tau_{k+1}> C\tilde{\Delta}[k]\}\bigg|\tilde{\Delta}[k]\bigg]\bigg].
\end{align}
We evaluate the inner expectations as 
\begin{align}
    \mathbb{E}_{\tau_{k+1}}\bigg[C\tilde{\Delta}[k]\mathbbm{1}\{\tau_{k+1}> C\tilde{\Delta}[k]\}\bigg|\tilde{\Delta}[k]\bigg]=C\tilde{\Delta}[k]\mathbb{P}(\tau_{k+1}> C\tilde{\Delta}[k])=C\tilde{\Delta}[k]e^{-\lambda_eC\tilde{\Delta}[k]},
\end{align}
and
\begin{align}
    \mathbb{E}_{\tau_{k+1}}\left[\tau_{k+1}\mathbbm{1}\{\tau_{k+1}> C\tilde{\Delta}[k]\}\bigg|\tilde{\Delta}[k]\right]&=\int_{C\tilde{\Delta}[k]}^{\infty}\tau_{k+1}\lambda_e e^{-\lambda_e\tau_{k+1}}d\tau_{k+1}\\
    &=\frac{1}{\lambda_e}e^{-\lambda_e C\tilde{\Delta}[k]}(\lambda_e C\tilde{\Delta}[k] + 1).
\end{align}
Therefore, we rewrite \eqref{avg_age3} as
\begin{align}
    \mathbb{E}\left[\beta_i[k]\right]\leq&\mathbb{E}_{\tilde{\Delta}[k]}\bigg[\underbrace{\frac{a^{(s)}[k]}{\lambda_e}\left(1-e^{-\lambda_e C\tilde{\Delta}[k]}(\lambda_e C\tilde{\Delta}[k] + 1)\right)}_{\text{bounded quantity}}+\underbrace{(a^{(s)}[k]-a^{(g)}[k])C\tilde{\Delta}[k]e^{-\lambda_eC\tilde{\Delta}[k]}}_{\text{bounded quantity}}\notag\\
    &+\underbrace{\frac{a^{(g)}[k]}{\lambda_e}e^{-\lambda_e C\tilde{\Delta}[k]}(\lambda_e C\tilde{\Delta}[k] + 1)}_{\text{bounded quantity}}\bigg].\label{bounded}
\end{align}

Now, we evaluate the asymptotic scaling $\lim_{n\to\infty}\mathbb{E}\left[\beta_i[k]\right]$. Since all the age metrics $a^{(s)}[k]$, $a^{(g)}[k]$ and $\tilde{\Delta}[k]$ on the right hand side of \eqref{bounded} are upper bounded by $k$ for all values of $n$, and $C$ is bounded, we use the bounded convergence theorem to exchange the limit and expectation to calculate its scaling as $n$ becomes large. From \eqref{a_k}, it is evident that $a^{(s)}[k]\sim O(1)$. From Lemma~\ref{lemma1}, we have that $a^{(g)}[k]\sim O(1)$. As $n\to\infty,$ the quantity $\lambda_{e}C\tilde{\Delta}[k]\to 0$. Thus, we have
\begin{align}
    \lim_{n\to\infty}\mathbb{E}\left[\beta_i[k]\right]\leq\lim_{n\to\infty}\frac{a^{(g)}[k]}{\lambda_e}=\lim_{n\to\infty}\frac{\lambda_{e}+\frac{B}{n-1}\tilde{a}[k]}{\lambda_e\left(\frac{\lambda}{n}+\frac{B}{n-1}\right)}=\frac{\lambda_{e}+\lambda \tilde{a}[k]}{\lambda_e\lambda}.\label{scaling}
\end{align}
Hence, using \eqref{scaling} in \eqref{avg_age2}, we obtain the asymptotic upper-bound for the average version age as
\begin{align}
    \lim_{n\to\infty}a_i\leq\lim_{k\to\infty}\lambda_e\times\frac{\lambda_{e}+\lambda \tilde{a}[k]}{\lambda_e\lambda}=2\frac{\lambda_{e}}{\lambda}+1,\label{final_UB}
\end{align}
concluding the proof.
\end{Proof}

\section{Average Age Scaling for Partial Connectivity}

So far, we have considered that all nodes are fully-connected to all other $n-1$ nodes of the network and can gossip without any restrictions. In a practical scenario, a single node can communicate properly with only a few number of nodes. To take this into account, we modify our system model, such that, each node can communicate only to a fraction $q$ of the $n-1$ nodes. In this new model, after the age sensing phase, when the network enters into the gossiping phase, each of the minimum-age nodes gossips only to $\lfloor q(n-1) \rfloor$ nodes chosen randomly.

To prove results about this system, we first consider a modified ASUMAN scheme, where the minimum-age nodes only gossip the data they have at the beginning of the interval $\mathcal{I}_k$. Clearly, such a system will have higher average age for a node, since in the interval of $\mathcal{I}_k$, the age of a gossiping node can only decrease due to updates from the source. Now, to formulate the age scaling of this modified scheme, we need some results about the statistical parameters of the minimum-age nodes. In the following lemma, we derive a lower bound for the probability of node $i$ not being a minimum-age node in the steady-state. 

\begin{lemma}\label{lemma3}
In a gossiping scheme, where in $\mathcal{I}_k$, the minimum-age nodes only gossip the data they have at time $T_k$, the steady-state probability of node $i$ not being a minimum age node satisfies the lower bound
\begin{align}
    \lim_{k\to\infty}\mathbb{P}(i\notin \mathcal{M}_k)\geq\frac{\lambda_e(\lambda-\lambda/n)}{\lambda_e^2+2\lambda_e\lambda+\lambda^2/n}.
\end{align}
\end{lemma}

\begin{figure}
\centerline{\includegraphics[width=0.5\textwidth]{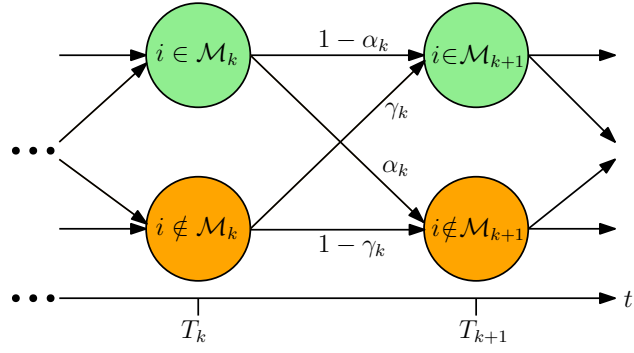}}
\caption{Probabilistic state transition of the random process.}
\label{MC}
\vspace*{-0.4cm}
\end{figure}

\begin{Proof}
At any $k$, node $i$ can either be in $\mathcal{M}_k$ or not. Depending on the updates it receives from the source and its gossiping neighbors, in the next time-stamp $k+1$, the node can be either in $\mathcal{M}_{k+1}$ or not. Given $i\in\mathcal{M}_k$, the event $i\in\mathcal{M}_{k+1}$ can only happen if either the source does not update any of the nodes in $\mathcal{I}_k$ or the source sends at least one update to node $i$ in $\mathcal{I}_k$. Hence, the probability of the event is $e^{-\lambda \tau_{k+1}}+(1-e^{-\frac{\lambda}{n}\tau_{k+1}})$. We denote the probability of the complementary event as $\alpha_k=e^{-\frac{\lambda}{n}\tau_{k+1}}-e^{-\lambda \tau_{k+1}}$ and the probability that given $i\notin\mathcal{M}_k$, $i\in\mathcal{M}_{k+1}$ as $\gamma_k$. This probabilistic transition is depicted in Fig.~\ref{MC}. The corresponding transition probabilities $(\alpha_k,\gamma_k)$, $k=1,2,3,\ldots$ are i.i.d., since the Poisson arrivals are memoryless. Now, we consider the random process $\mathbbm{1}\{i\notin \mathcal{M}_k\}$. The evolution of the process can be expressed by the following recurrence relation:
\begin{align}
    \mathbbm{1}\{i\notin \mathcal{M}_{k+1}\}=
\left\{
	\begin{array}{lll}
		1, &1-\gamma_k, & \mathbbm{1}\{i\notin \mathcal{M}_k\}=1\\
		1, &\alpha_k, & \mathbbm{1}\{i\notin \mathcal{M}_k\}=0\\
        0, &\text{otherwise.}&
	\end{array}
\right.
\end{align}
Therefore, we can write the expectation of the process as
\begin{align}\label{MCiid1}
    \mathbb{E}[\mathbbm{1}\{i\notin \mathcal{M}_{k+1}\}]=\mathbb{P}(i\notin \mathcal{M}_{k+1})=\mathbb{E}[1-\gamma_k]\mathbb{P}(i\notin \mathcal{M}_{k})+\mathbb{E}[\alpha_k]\mathbb{P}(i\in \mathcal{M}_{k}).
\end{align}
Similarly, we can write the expectation of the process $\mathbbm{1}\{i\in \mathcal{M}_{k+1}\}$ as
\begin{align}\label{MCiid2}
    \mathbb{E}[\mathbbm{1}\{i\in \mathcal{M}_{k+1}\}]=\mathbb{P}(i\in \mathcal{M}_{k+1})=\mathbb{E}[\gamma_k]\mathbb{P}(i\notin \mathcal{M}_{k})+\mathbb{E}[1-\alpha_k]\mathbb{P}(i\in \mathcal{M}_{k}).
\end{align}
Combining \eqref{MCiid1} and \eqref{MCiid2}, we can write
\begin{align}
    [\mathbb{P}(i\notin \mathcal{M}_{k+1}) \quad \mathbb{P}(i\in \mathcal{M}_{k+1})]=[\mathbb{P}(i\notin \mathcal{M}_{k}) \quad \mathbb{P}(i\in \mathcal{M}_{k})]\times P,
\end{align}
where $P$ is the transition matrix
\begin{align}\label{MCiid3}
    P=\left[ \begin{array} {cc} 1-\mathbb{E}[\gamma_k] & \mathbb{E}[\gamma_k] \\ \mathbb{E}[\alpha_k]& 1-\mathbb{E}[\alpha_k] \\ \end{array} \right].
\end{align}
Clearly, \eqref{MCiid3} is a time-homogeneous Markov chain equation, and we can determine the stationary distribution of such a process as
\begin{align}\label{stationary_prob}
    \lim_{k\to\infty}\mathbb{P}(i\notin \mathcal{M}_k)=\frac{\mathbb{E}[\alpha_k]}{\mathbb{E}[\alpha_k]+\mathbb{E}[\gamma_k]},
\end{align}
where the right hand side does not depend on $k$ due to the i.i.d.~nature of $(\alpha_k,\gamma_k)$. We can calculate $\mathbb{E}[\alpha_k]= \mathbb{E}\left[e^{-\frac{\lambda}{n}\tau_{k+1}}-e^{-\lambda \tau_{k+1}}\right]=\frac{\lambda_e}{\lambda_e+{\lambda}/{n}}-\frac{\lambda_e}{\lambda_e+\lambda}=\frac{\lambda_e(\lambda-\lambda/n)}{(\lambda_e+\lambda/n)(\lambda_e+\lambda)}$. On the other hand, from \eqref{stationary_prob}, it is clear that the stationary distribution decreases in $\mathbb{E}[\gamma_k]$. Since $\gamma_k$ is a probability, the maximum value of $\mathbb{E}[\gamma_k]$ can be $1$, and we can write the following lower bound 
\begin{align}
    \lim_{k\to\infty}\mathbb{P}(i\notin \mathcal{M}_k)\geq\frac{\frac{\lambda_e(\lambda-\lambda/n)}{(\lambda_e+\lambda/n)(\lambda_e+\lambda)}}{\frac{\lambda_e(\lambda-\lambda/n)}{(\lambda_e+\lambda/n)(\lambda_e+\lambda)}+1}=\frac{\lambda_e(\lambda-\lambda/n)}{\lambda_e^2+2\lambda_e\lambda+\lambda^2/n}.
\end{align}
completing the proof.
\end{Proof}

Now, using the result of Lemma~\ref{lemma3}, we formulate the average version age of a node in Theorem~\ref{thm2}.

\begin{theorem}\label{thm2}
Under the same conditions as Theorem~\ref{thm1}, i.e., $C\to 0$ as $n\to\infty$ and $B=n\lambda$, the average version age of a node in a network with partial connectivity scales as $O(1)$.
\end{theorem}

\begin{Proof}
We prove this theorem in the same way as the first part of Theorem~\ref{thm1}. We define a new process $\Delta^{(p)}[k]$ under the modified ASUMAN scheme, such that $\Delta^{(p)}[1]=1$. The process evolves as follows: $\Delta^{(p)}[k]=\tilde{\Delta}[k-1]+1$ only if $i\notin\mathcal{M}_{k-1}$ and it gets at least one update from some other node in $\mathcal{M}_{k-1}$. Otherwise,  $\Delta^{(p)}[k]=\Delta^{(p)}[k-1]+1$. We denote $N_i(\tau_{k+1})=\sum_{j\in\mathcal{M}_k}N_{i;j}(\tau_{k+1})$ as the total number of gossip updates received by the $i$th node from $j$ in $\mathcal{I}_k$. Using these notations, we can express the probabilistic recurrence relations for $k>1$ as
\begin{align}
    \Delta^{(p)}[k]=\left\{
	\begin{array}{ll}
		\tilde{\Delta}[k-1]+1, &\pi(\tau_k)\\
		\Delta^{(p)}[k-1]+1, &1-\pi(\tau_k)
	\end{array}
\right.\label{delta_s2}
\end{align}
where
\begin{align}
    \pi(\tau_k)=\mathbb{P}(\tau_k > C\tilde{\Delta}[k-1])\mathbb{P}(i\notin\mathcal{M}_{k-1})\mathbb{P}(N_i(\tau_{k})\geq 1).\label{pi_tau_k}
\end{align}
Clearly, $\Delta_i(t)\leq\Delta^{(p)}[k]$ for $t\in\mathcal{I}_k$. We evaluate the mean of this process as
\begin{align}
    a^{(p)}[k]&= \mathbb{E}[\Delta^{(p)}[k]]\\
    &= \mathbb{E}\big[(\tilde{\Delta}[k-1]+1)\pi(\tau_k)+(\Delta^{(p)}[k-1]+1)(1-\pi(\tau_k))\big]\\
    &= \mathbb{E}\big[(\tilde{\Delta}[k-1]+1)\pi(\tau_k)\big]+(a^{(p)}[k-1]+1)(1-\mathbb{E}[\pi(\tau_k)]).\label{partialExpt}
\end{align}
Now, to obtain the asymptotic age scaling, we evaluate the first expectation in \eqref{partialExpt} as 
\begin{align}      
    \lim_{n\to\infty}\mathbb{E}&\left[(\tilde{\Delta}[k-1]+1)\pi(\tau_k)\right]\notag\\
    &=\mathbb{E}\left[\lim_{n\to\infty}\left(\left(\tilde{\Delta}[k-1]+1\right)\mathbb{P}(\tau_k > C\tilde{\Delta}[k-1])\mathbb{P}(i\notin\mathcal{M}_{k-1}) \mathbb{P}(N_i(\tau_{k})\geq 1)\right)\right],\label{expt_tau}
\end{align}
since the terms inside the expectations are bounded quantities. From the proof of Theorem~\ref{thm1}, we know that $\lim_{n\to\infty}\mathbb{P}(\tau_k > C\tilde{\Delta}[k-1])=1.$ From Lemma~\ref{lemma3}, $\lim_{n\to\infty}\mathbb{P}(i\notin\mathcal{M}_{k-1})\geq\lambda/(\lambda_e+2\lambda)$. We evaluate the third probability in \eqref{expt_tau} as 
\begin{align}
    \mathbb{P}(N_i(\tau_{k})\geq 1)
    =1-\mathbb{P}(N_i(\tau_{k})=0)=1-\prod_{j=1}^{|\mathcal{M}_k|}\mathbb{P}(N_{i;j}(\tau_{k})=0).\label{expt_n_j}
\end{align}
In our system model, node $i$ can receive no updates from node $j$, if either $j$ is not connected to $i$, or it is connected to $i$ but does not send any updates. The probability of this event is $(1-q)+qe^{-\frac{B(\tau_k-C\tilde{\Delta}[k-1])}{|\mathcal{M}_k|\lfloor q(n-1)\rfloor}}$.
Therefore, we rewrite \eqref{expt_n_j} as 
\begin{align}
    \mathbb{P}(N_i(\tau_{k})\geq 1)=1-\left((1-q)+qe^{-\frac{B(\tau_k-C\tilde{\Delta}[k-1])}{|\mathcal{M}_k|\lfloor q(n-1)\rfloor}}\right)^{|\mathcal{M}_k|}.\label{expt_n_exact}
\end{align}
Since \eqref{expt_n_exact} is an increasing function of $|\mathcal{M}_k|\geq 1$, it attains the minimum value for $|\mathcal{M}_k|=1$. We evaluate the expression as the following lower bound
\begin{align}
    \mathbb{P}(N_i(\tau_{k})\geq 1)\geq 1-\left((1-q)+qe^{-\frac{B(\tau_k-C\tilde{\Delta}[k-1])}{\lfloor q(n-1)\rfloor}}\right)=q\left(1-e^{-\frac{B(\tau_k-C\tilde{\Delta}[k-1])}{\lfloor q(n-1)\rfloor}}\right).\label{expt_n}
\end{align}
Now, applying this inequality along with the relation in $B=n\lambda$ in \eqref{expt_n}, we obtain
\begin{align}
    \mathbb{E}\left[ \lim_{n\to\infty}\mathbb{P}(N_i(\tau_{k})\geq 1)\right]\geq\mathbb{E}\left[q\left(1-e^{-\frac{\lambda}{q}\tau_k}\right)\right]=\frac{q\lambda}{\lambda+q\lambda_e}.\label{expt_p_n}
\end{align}
Using \eqref{expt_p_n} in \eqref{pi_tau_k}, we get the following lower bound
\begin{align}
    \lim_{n\to\infty}\mathbb{E}[\pi(\tau_k)]\geq\frac{q\lambda^2}{(\lambda_e+2\lambda)(\lambda+q\lambda_e)}.\label{expt_pi}
\end{align}

Let us denote the quantity on the right-hand side of \eqref{expt_pi}, as $\tilde{\pi}$. Clearly, $\lim_{n\to\infty}\mathbb{E}[\pi(\tau_k)]\geq\tilde{\pi}$, and thus, $1-\lim_{n\to\infty}\mathbb{E}[\pi(\tau_k)]\leq 1-\tilde{\pi}$. Now, since $a^{(p)}[k]\geq\tilde{a}[k]\ \forall k$, we can write the following
\begin{align}
    \lim_{n\to\infty}a^{(p)}[k]&\leq\left(\lim_{n\to\infty}\tilde{a}[k-1]+1\right)\tilde{\pi}+\left(\lim_{n\to\infty}a^{(p)}[k-1]+1\right)(1-\tilde{\pi})\\
    &\leq\left(2+\frac{\lambda_e}{\lambda}\right)\tilde{\pi}+\left(\lim_{n\to\infty}a^{(p)}[k-1]+1\right)(1-\tilde{\pi}),\label{expt_final}
\end{align}
where \eqref{expt_final} follows from Lemma~\ref{lemma1}. Now, using the similar logic of recursive inequalities as in Theorem~\ref{thm1}, we can write that $\lim_{n\to\infty}a^{(p)}\leq 1+\frac{\lambda_e}{\lambda}+\frac{1}{\tilde{\pi}}\sim O(1)$, thus completing the proof.
\end{Proof}

\section{Average Age Scaling for Networks with Finite Connectivity}

In the previous section, we have considered a network with partial, nevertheless $O(n)$, connectivity, and the age scaling turned out to be $O(1)$. In this section, we investigate the $O(1)$ connectivity networks and derive results for their age scaling in the settings of ASUMAN.

\begin{figure}[t]
\subfigure[ring network]{
\centering
\includegraphics[width=0.4\textwidth]{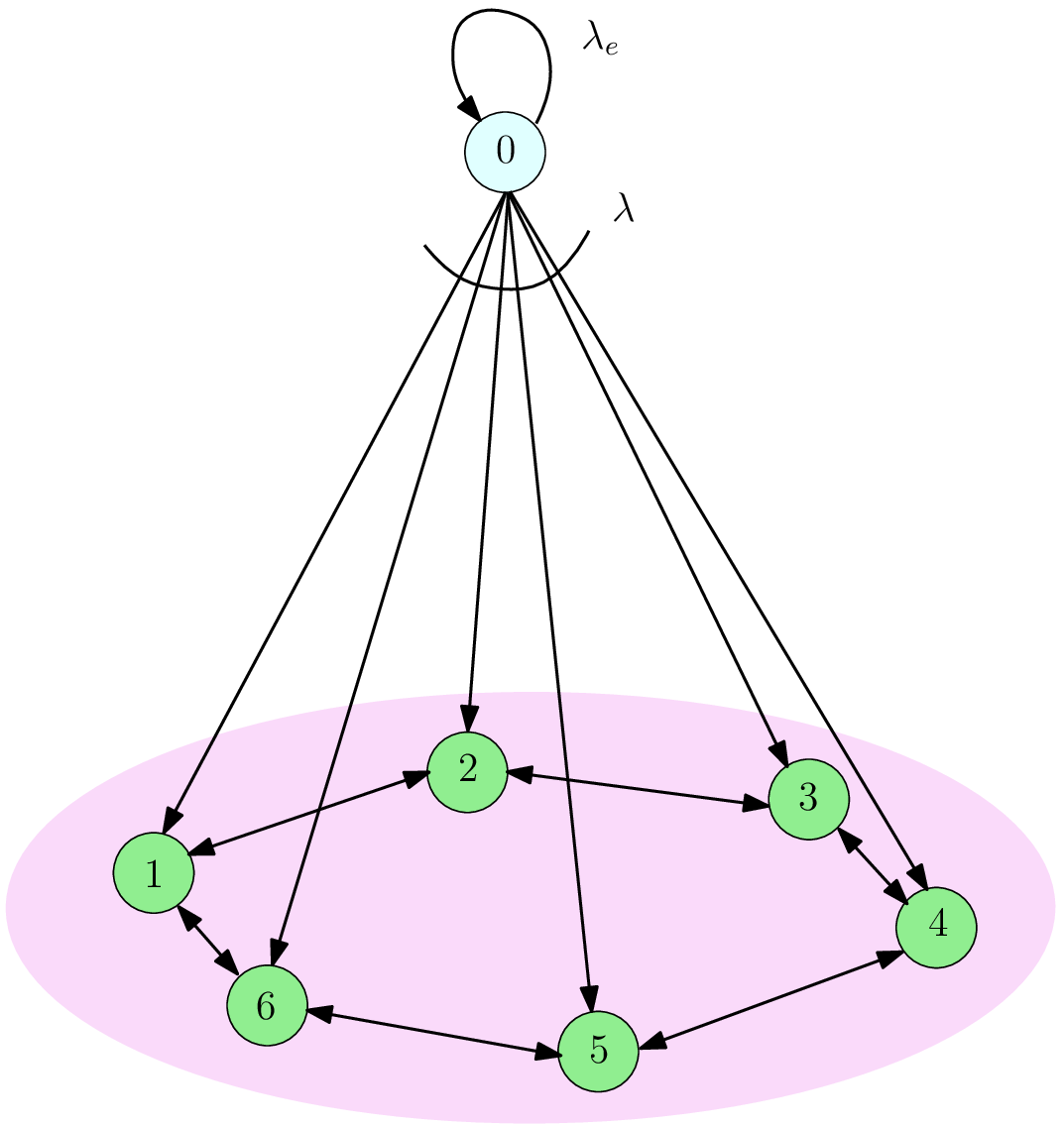}
\label{ring_pic}
}\hfill
\subfigure[two-dimensional grid network]{
\centering
\includegraphics[width=0.4\textwidth]{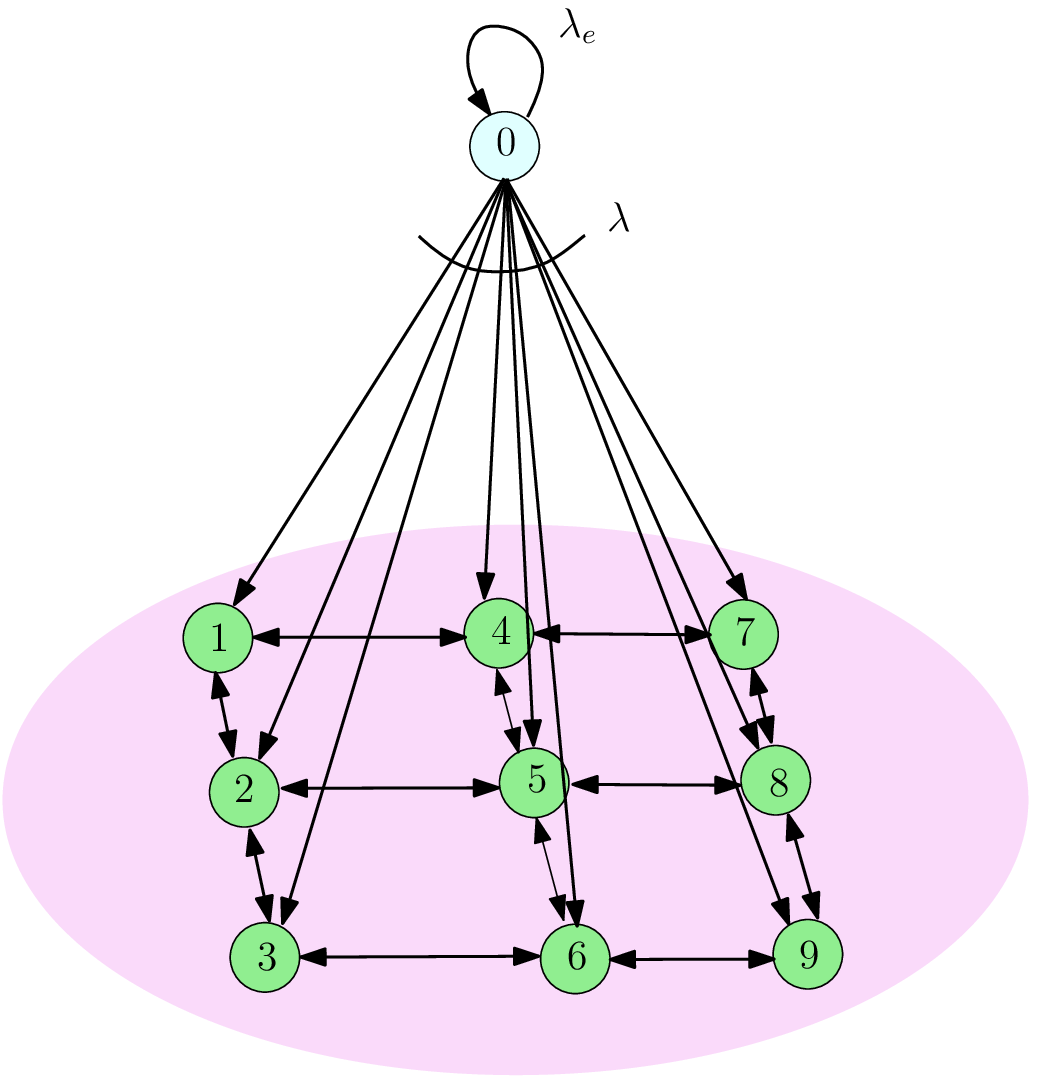}
\label{2D_grid_pic}
}
\caption{Network structures with finite connectivity. In (a), each node is connected to two neighbors on a ring network, and in (b) each node is connected to four neighbors on a grid network.}
\vspace*{-0.4cm}
\end{figure}

\begin{theorem}\label{finite_connectivity}
For any symmetric network with $n$ nodes, where each node is connected to $O(1)$ neighboring nodes, the age of any individual node scales as $\Omega(n)$.
\end{theorem}

\begin{Proof}
First, we consider a bidirectional ring network, as shown in Fig.~\ref{ring_pic}, where each node can gossip to two of its neighboring nodes. Clearly, in the settings of ASUMAN, there will be gossip in a node when either a node is updated by the source itself or any of its two neighbors is a minimum age node. Now, consider the process $\Delta^{(r)}[k]$: It increases by $1$, when node $i$ and its neighbors do not receive any update from the source in $I_k$ and it decreases to $0$ otherwise. We initialize $\Delta^{(r)}[1]=1$. Therefore, for $k>1$, we have the following recurrence relation
\begin{align}
        \Delta^{(r)}[k]=
\left\{
	\begin{array}{ll}
		\Delta^{(r)}[k-1]+1, &e^{-\frac{3\lambda}{n}\tau_k}\\
		0, &1-e^{-\frac{3\lambda}{n}\tau_k}.
	\end{array}
\right.
\end{align}
Clearly, $\Delta^{(r)}[k]\leq\Delta_i(t)$ for $t\in\mathcal{I}_k$. We evaluate the mean of this process as
\begin{align}
    a^{(r)}[k]=\mathbb{E}[\Delta^{(r)}[k]]=a^{(r)}[k-1]\mathbb{E}[e^{-\frac{3\lambda}{n}\tau_k}]=a^{(r)}[k-1]\left(\frac{\lambda_e}{\lambda_e+\frac{3\lambda}{n}}\right).
\end{align}
Now, evaluating the steady state mean as in Lemma~\ref{lemma1}, we obtain
\begin{align}
    a^{(r)}=\lim_{k\to\infty}a^{(r)}[k]=\frac{\frac{\lambda_e}{\lambda_e+\frac{3\lambda}{n}}}{1-\frac{\lambda_e}{\lambda_e+\frac{3\lambda}{n}}}=\frac{n\lambda_e}{3\lambda}.
\end{align}
Now, using the relation $a_i\geq a^{(r)}$, we conclude $a_i=\Omega(n)$. We can do similar calculations for other $O(1)$ connected networks, such as, the two-dimensional grid, the three-dimensional cube, etc., and reach the same conclusion. This completes the proof. 
\end{Proof}

The analysis in \cite{buyukates21CommunityStructure,buyukates22ClusterGossip} shows that uniform gossip in a ring network yields $O(\sqrt{n})$ age. Therefore, we conclude that uniform gossip performs better than ASUMAN for ring networks.

\section{ASUMAN with Sub-Linear Connectivity} \label{sect:sub-linear}

\begin{figure}[t]
\subfigure[disconnected cluster heads]{
\centering
\includegraphics[width=0.3\textwidth]{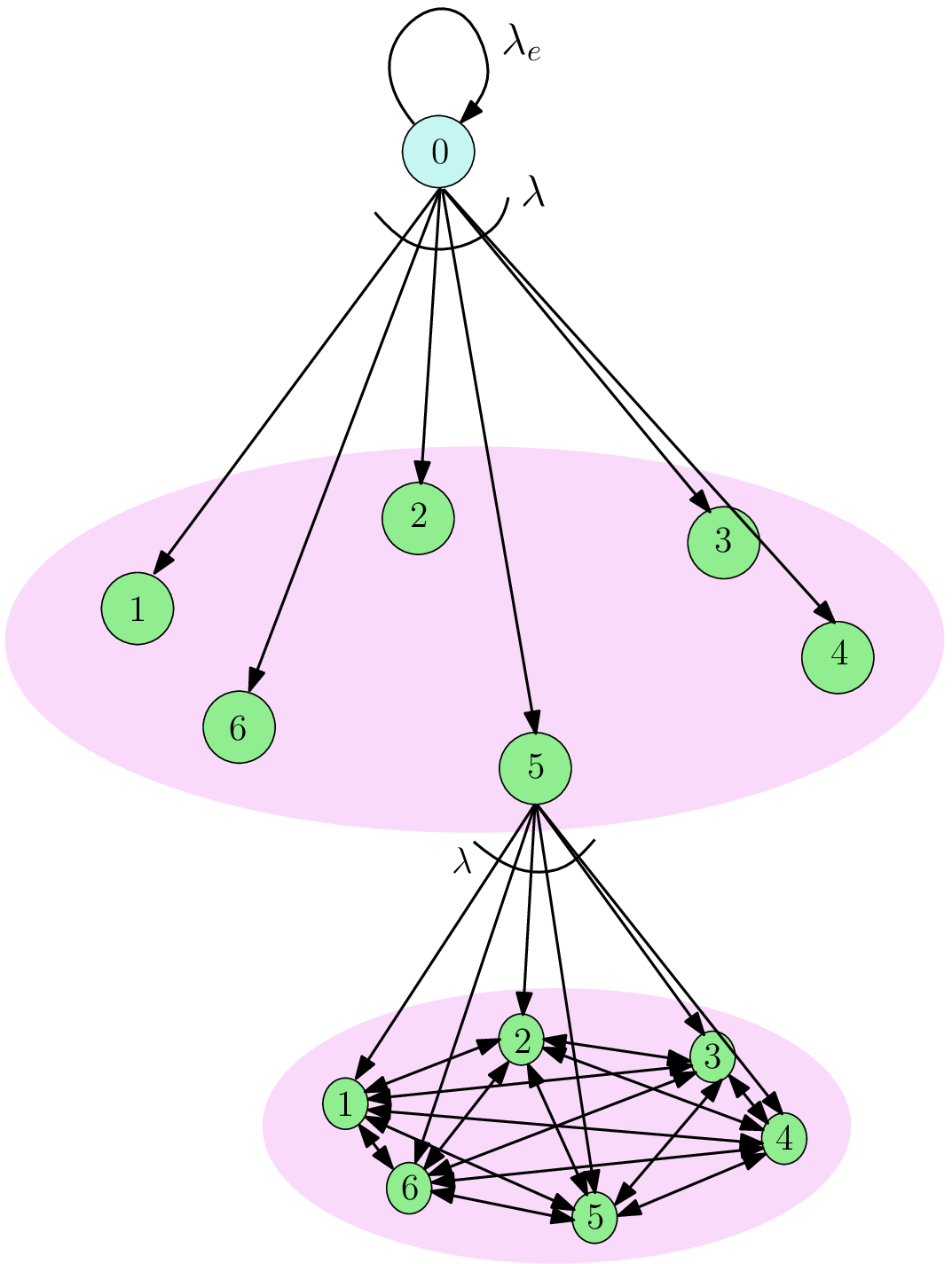}
\label{disC}
}\hfill
\subfigure[ring-connected cluster heads]{
\centering
\includegraphics[width=0.3\textwidth]{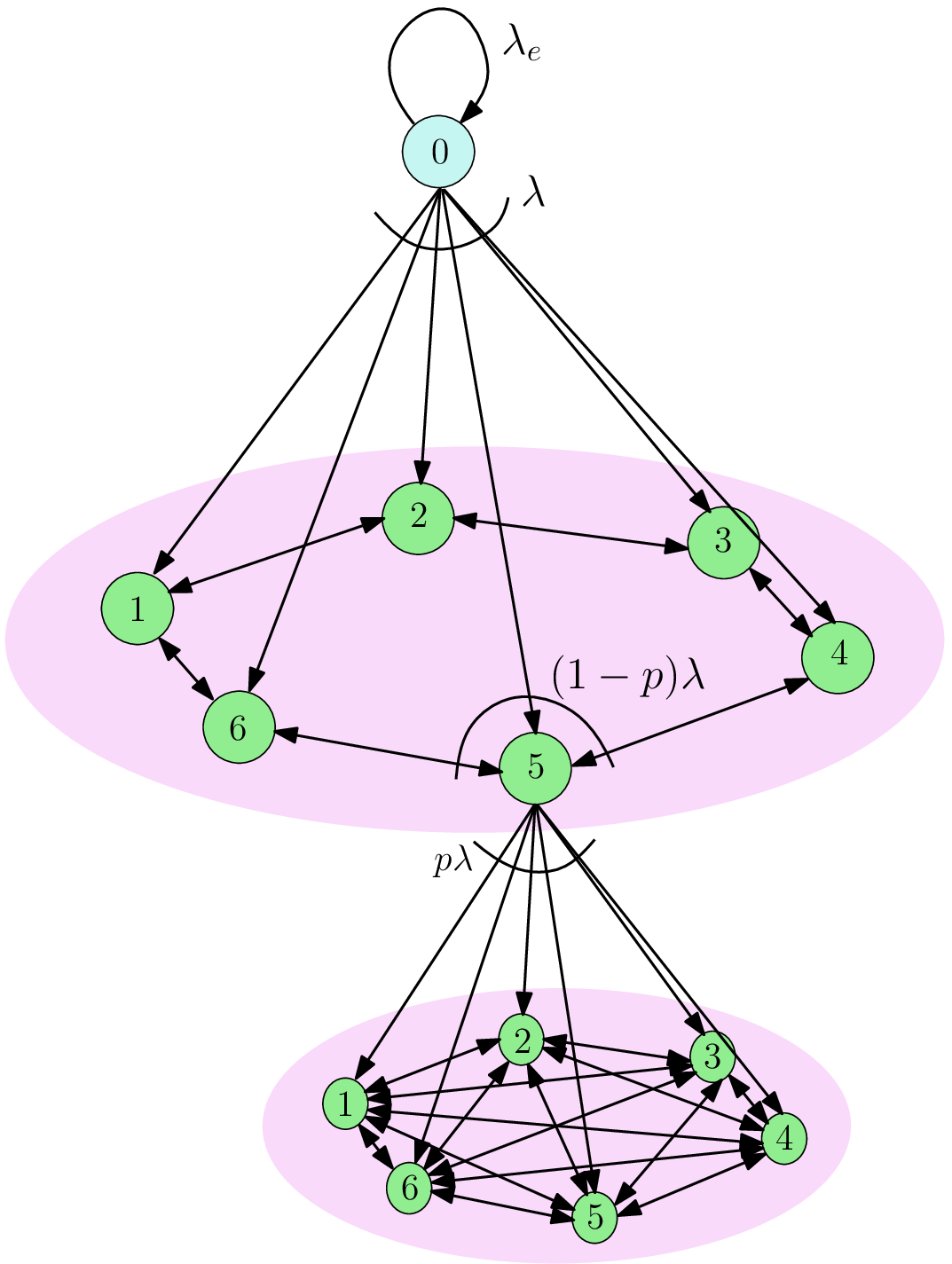}
\label{ringC}
}\hfill
\subfigure[fully-connected cluster heads]{
\centering
\includegraphics[width=0.3\textwidth]{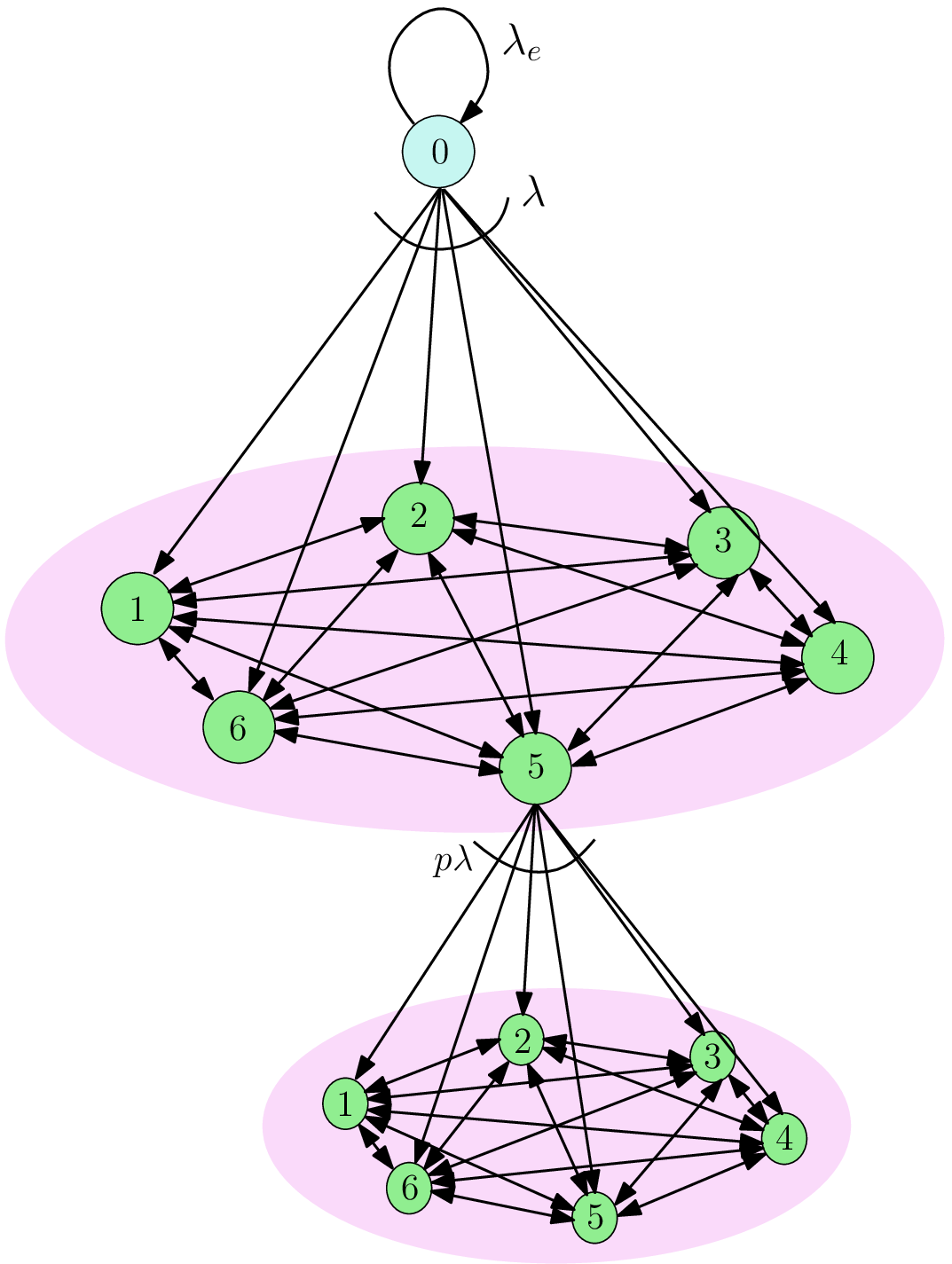}
\label{fullC}
}
\caption{Clustered gossip networks with different topologies of cluster heads: In (a), cluster heads are disconnected (i.e., do nor gossip), in (b) cluster heads are connected on a ring, and in (c) cluster heads are fully-connected.}
\label{ClusteredDiagram}
\vspace*{-0.4cm}
\end{figure}

In the previous two sections, we have observed that $O(n)$ connectivity yields $O(1)$ age scaling, whereas $O(1)$ connectivity yields an $\Omega(n)$ age scaling, for ASUMAN. To find a trade-off between these two settings, we modify our assumption of a single layer fully-connected network and introduce a hierarchical structure. We divide the $n$ nodes into $c$ clusters, each containing $m$ nodes, such that $n=c\cdot m$. From each cluster, one node is selected as the cluster head. Those heads form a fully-connected network of $c$ nodes. Each node in a cluster is connected to $m-1$ neighboring nodes, and each cluster head is connected to $c+m-2$ nodes. Therefore, this system model, shown in Fig.~\ref{fullC}, ensures that the connectivity of each node is of the order $O(\max(c+m,m))$. Every  cluster head receives updates from the source with Poisson arrival rate $\frac{\lambda}{c}$ and has a reserve of $p\lambda$ update rate for updating its nodes ($0\leq p \leq 1$) and $(1-p)\lambda$ update rate for gossiping with the other cluster heads. The cluster nodes use their available update rate to gossip locally (within each cluster) with ASUMAN. We denote such a cluster as $\mathcal{C}$ and formulate the age of an individual leaf node.

\begin{theorem}
For a dense cluster network with fully-connected cluster heads, the upper bound on the average age of a cluster node scales as $O(1)$ with $O(\sqrt{n})$ connectivity. This upper bound is minimized with the optimal choice $p^*=\frac{1}{2}$.    
\end{theorem}

\begin{Proof}
Choosing $c=\sqrt{n}$, we get $O(\sqrt{n})$ connectivity. To analyze this hierarchical network, we modify the SHS formulation for getting an upper bound on the average age. The age of each node in the cluster increases by $1$, when the source updates itself with rate $\lambda_e$. The nodes gossip in the same opportunistic manner within the cluster nodes; thus, only the minimum age nodes gossip in any $\mathcal{I}_k$ time slot and upon receiving an update a node updates its information to the least old version. The only difference in this formulation from our original SHS formulation in Section~\ref{sect:asuman} is the change in direct updates. Since in this model a cluster node is directly updated from only a cluster head with rate $\frac{p\lambda}{c}$ and not by the actual source, such transitions make the age of the cluster node $\Delta_{(1)}(t)$, the current age of the cluster head.

Under these modifications, the extended generator function for the $i$th cluster node, with the same test function $\psi_i$, can be written as
\begin{align}
    \mathbb{E}[(L\psi_{i})(\mathbf{\Delta}_{(2)}(t),t)]
    =&\mathbb{E}\bigg[\lambda_{e}(\Delta_i(t)+1-\Delta_i(t))+\frac{\lambda}{h}(\Delta_{(1)}(t)-\Delta_i(t)) \nonumber\\
    & \qquad +\sum_{j\in \mathcal{C}}\lambda_{j,i}(\mathbf{\Delta}(t),t)\left(\Delta_{\{j,i\}}(t)-\Delta_i(t)\right)\bigg]=0.\label{cluster_eqn}
\end{align}
Now, using the same argument as in Section~\ref{sect:asuman}, we obtain the upper bound $\Delta_{\{j,i\}}(t)\leq\tilde{\Delta}_{(2)}[k]$ for $t\in\mathcal{I}_k$, where $\tilde{\Delta}_{(2)}[k]$ is the minimum age process of the cluster. We can write this process of the cluster as the following recursive relation
\begin{align}
    \tilde{\Delta}_{(2)}[k+1]=
\left\{
	\begin{array}{ll}
		\tilde{\Delta}_{(2)}[k]+1, &e^{-p\lambda\tau_k}\\
		\Delta_{(1)}[k]+1, &1-e^{-p\lambda\tau_k}
	\end{array}
\right.
\end{align}
Now, solving this using recursive expectations as in Lemma~\ref{lemma1}, we obtain
\begin{align}
    \tilde{a}_{(2)}=\lim_{k\to\infty}\tilde{a}_{2}[k]=\frac{\lambda_e}{p\lambda}+a_{(1)}+1.\label{cluster_min_age}
\end{align}
Using \eqref{cluster_min_age} in the expression of \eqref{cluster_eqn}, we get the following SHS equation
\begin{align}
    a_{(2)}=\lim_{t\to\infty}\mathbb{E}[\Delta_{i}(t)]\leq\frac{\lambda_e+\frac{p\lambda}{m}a_{(1)}+\frac{m\lambda}{m-1}\left(\frac{\lambda_e}{p\lambda}+a_{(1)}+1\right)}{\frac{p\lambda}{m}+\frac{m\lambda}{m-1}}. \label{Hierarchical_age}
\end{align}
The cluster heads are connected by ASUMAN scheme with total update rate $B=c(1-p)\lambda$. Using the formula in \eqref{UB2}, we obtain
\begin{align}
    a_{(1)}\leq\frac{\lambda_e}{\lambda}\frac{1+\frac{c(1-p)\lambda}{c-1}\left(\frac{1}{\lambda}+\frac{1}{\lambda_e}\right)}{\frac{1}{c}+\frac{c(1-p)}{c-1}}.\label{UB_fc}
\end{align}
Since the right hand side in \eqref{UB_fc} is an increasing function of $c$, taking $c=\sqrt{n}\to\infty$, we obtain a simpler upper bound $a_{(1)}\leq(1+\frac{1}{1-p})\frac{\lambda_e}{\lambda}+1$. Substituting this upper bound of $a_{(1)}$ and taking $m=\sqrt{n}\to\infty$, we obtain
\begin{align}
    \lim_{m\to\infty}a_{(2)}\leq\left(2+\frac{1}{p}+\frac{1}{1-p}\right)\frac{\lambda_e}{\lambda}+2 \sim O(1),
\end{align}
which concludes the first part of the proof.

To minimize this upper bound, we minimize the convex function $\frac{1}{p}+\frac{1}{1-p}$, where $p\in [0,1]$. Using KKT conditions, the optimum value of $p$ is  $p^*=\frac{1}{2}$, resulting $a_{(2)}^*=\frac{6\lambda_e}{\lambda}+2$.
\end{Proof}

We observe that this reduction in connectivity, while keeping the age performance the same, i.e., $O(1)$, is possible due to the hierarchical structure of the network which allows a cluster of nodes to track the age of their cluster head, which was not possible previously due to every node tracking only the source. Also, this setting enables some nodes (cluster heads) to have lower age than others (cluster nodes).

\section{Average Age Scaling for Clustered Network Topologies}

The hierarchical structure, introduced in Section~\ref{sect:sub-linear}, yields lower order of connectivity while maintaining $O(1)$ age performance. This is similar to the analysis of \cite{buyukates21CommunityStructure,buyukates22ClusterGossip}, where clustering brings down the order of age scaling. However, in practical scenarios, often the first layer of the hierarchy is not a fully-connected network. In this section, we investigate the effect of connectivity of cluster heads. From Section~\ref{sect:sub-linear}, we already know that for partial connectivity of cluster heads we will get $O(1)$ age performance since $a_{(1)}\sim O(1)$. Now, we look into two more structures, namely, disconnected cluster heads (see Fig.~\ref{disC}) and ring-connected cluster heads (see Fig.~\ref{ringC}), similar to the work in \cite{buyukates21CommunityStructure,buyukates22ClusterGossip}, and formulate their age performance.

\begin{theorem}
For a hierarchical cluster network with disconnected and ring-connected cluster heads, the upper bound of average age of a cluster node scales as $O(c)$ and $O(\sqrt{c})$, respectively.
\end{theorem}

\begin{Proof}
For disconnected cluster heads, $p=1$ and the analysis in \cite{yates21gossip}, shows $a_{(1)}=c\frac{\lambda_{e}}{\lambda}$. Substituting in \eqref{Hierarchical_age}, we get 
\begin{align}
    \lim_{m\to\infty}a_{(2)}\leq(2+c)\frac{\lambda_e}{\lambda}+1\sim O(c).
\end{align}
Similarly, from the analysis of \cite{buyukates22ClusterGossip}, we obtain that the average age of a ring-connected cluster head is $a_{(1)}\approx\sqrt{\frac{\pi}{2}}\frac{\lambda_e}{\lambda}\sqrt{\frac{c}{1-p}}$. Substituting in \eqref{Hierarchical_age}, we get 
\begin{align}
    \lim_{m\to\infty}a_{(2)}\leq\left(1+\frac{1}{p}+\sqrt{\frac{\pi c}{2(1-p)}}\right)\frac{\lambda_e}{\lambda}+1 \sim O(\sqrt{c}), \label{p_ub}
\end{align}
concluding the proof.
\end{Proof}

\section{Average Age Scaling for Asymmetric Update Rates}\label{sect:asymmetric} 

\begin{figure}[t]
\centerline{\includegraphics[width=0.5\textwidth]{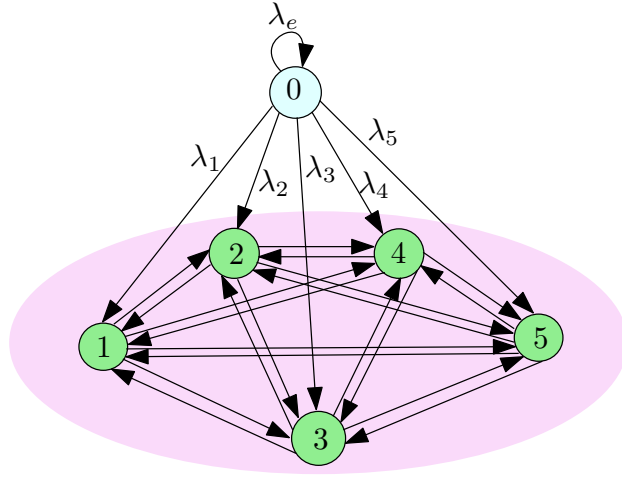}}
\caption{An asymmetric arrival gossip network, where the source updates the nodes with unequal update rates $\{\lambda_i\}_{i=1}^n$, i.e.,  $\lambda_i\neq \frac{\lambda}{n}$ here. However, the total update rate is still $\sum_{i=1}^{n}\lambda_i=\lambda$ as before.}
\label{asymm}
\vspace*{-0.4cm}
\end{figure}

So far, we have only considered symmetric networks. However, in practice, the network can be asymmetric. We model such an asymmetric network as shown in Fig.~\ref{asymm}. Here, all the $n$ nodes have different update rates $\{\lambda_i\}_{i=1}^n$ from the source. However, the source still has a total update rate constraint $\sum_{i=1}^{n}\lambda_i=\lambda$ as before. In the following theorem, we formulate the dependence of the upper bound of a node's average age and its update rate. Later in this section, we present an example of such an asymmetric network with power law arrivals.

\begin{theorem}\label{thm_asymm}
In an asymmetric network with ASUMAN gossiping scheme, the upper bound of the age of any node scales as $O(1)$. All the upper bounds converge to the same value, if for any $i\in\mathcal{N}$, $\lambda_i\to 0$ as $n\to\infty$. Further, the  upper bound can be minimized only up to $\frac{\lambda_e}{\lambda}+\frac{1}{2}$. 
\end{theorem}

\begin{Proof}
Using the formulation as in Lemma~\ref{lemma2}, we get the following result 
\begin{align}\label{age_asymm}
    a_i\leq\frac{\lambda_{e}}{\lambda}\frac{\left(1+\frac{B}{n-1}\left(\frac{1}{\lambda}+\frac{1}{\lambda_{e}}\right)\right)}{\left(\frac{\lambda_i}{\lambda}+\frac{n}{n-1}\right)}.
\end{align}
Since $0\leq\lambda_i\leq\lambda$, the asymptotic scaling becomes
\begin{align}
    \lim_{n\to\infty}a_i\leq\lim_{n\to\infty}\frac{\lambda_{e}}{\lambda}\frac{\left(1+\frac{n\lambda}{n-1}\left(\frac{1}{\lambda}+\frac{1}{\lambda_{e}}\right)\right)}{\left(\frac{n}{n-1}\right)}=2\frac{\lambda_e}{\lambda}+1.\label{asymm_eqn}
\end{align}

Clearly, the gossiping average age scales as $O(1)$, and the age sensing phase does not affect the age scaling. Therefore, even with asymmetric updates, average age scales as $O(1)$. Also, from \eqref{asymm_eqn}, it is clear that if $\frac{\lambda_i}{\lambda}\to 0$ as $n\to\infty$ for any $i\in\mathcal{N}$, then all the age scaling upper bounds converges to $2\frac{\lambda_e}{\lambda}+1$. This proves the first part of the theorem.

From the formulation in \eqref{asymm_eqn}, it is clear that if $\lim_{n\to\infty}\frac{\lambda_i}{\lambda}=1$, then the bound becomes
\begin{align}
    \lim_{n\to\infty}a_i\leq\frac{\lambda_e}{\lambda}+\frac{1}{2},
\end{align}
proving the second part of the theorem, and concluding the proof.
\end{Proof}

\subsection{Special Case: Power Law Arrivals}

In this example, the source-to-node update rates have the form $\lambda_i=\theta\nu^i$ with $\sum_{i=1}^{n}\lambda_i=\lambda$. Clearly, if $\nu=1$ then $\lambda_i=\frac{\lambda}{n}$. However, for $0<\nu< 1$, using the relation $\sum_{i=1}^{n}\lambda_i=\lambda$, we have $\theta=\frac{\lambda(1-\nu)}{\nu(1-\nu^n)}$ from our choice of $\nu$. Since for asymptotic behavior we only need $\frac{\lambda_i}{\lambda}$, using the expression of $\theta$, we obtain
\begin{align}\label{eta_formula}
    \frac{\lambda_i}{\lambda}=\frac{\nu^i}{1-\nu^n}\left(\frac{1-\nu}{\nu}\right),\qquad 0< \nu < 1.
\end{align}
Now, substituting \eqref{eta_formula} in \eqref{age_asymm}, we get the upper bound of the age of $i$th node as
\begin{align}\label{eta_age}
    a_i\leq\frac{\lambda_{e}}{\lambda}\frac{\left(1+\frac{n\lambda}{n-1}\left(\frac{1}{\lambda}+\frac{1}{\lambda_{e}}\right)\right)}{\left(\frac{\nu^i}{1-\nu^n}\left(\frac{1-\nu}{\nu}\right)+\frac{n}{n-1}\right)}.
\end{align}
For very large $n$, the upper bound of the age becomes
\begin{align}\label{eta_asymp}
    \lim_{n\to\infty}a_i\leq\lim_{n\to\infty}\frac{\lambda_{e}}{\lambda}\frac{\left(1+\frac{n\lambda}{n-1}\left(\frac{1}{\lambda}+\frac{1}{\lambda_{e}}\right)\right)}{\left(\frac{\nu^i}{1-\nu^n}\left(\frac{1-\nu}{\nu}\right)+\frac{n}{n-1}\right)}=\frac{\lambda_{e}}{\lambda}\frac{\left(2+\frac{\lambda}{\lambda_{e}}\right)}{\left(1+\nu^i\left(\frac{1-\nu}{\nu}\right)\right)}.
\end{align}
In \eqref{eta_asymp}, we notice that the upper bound of the age is much more tight for nodes with higher update rate from the source and loose for nodes with fewer updates. In fact, when $i\approx n\to \infty$, the upper bound is the maximum limit $2\frac{\lambda_e}{\lambda}+1$.

\section{Numerical Results}

In this section, we compare our analytically derived results with numerical simulations. We choose $C=\frac{1}{n}$ for the simulations and calculate the average version age of a single node for up to $n=600$ nodes. We use two different values for $\frac{\lambda_{e}}{\lambda}$, specifically, $\frac{\lambda_{e}}{\lambda}=1$ and $\frac{\lambda_{e}}{\lambda}=2$. We also simulate the average version age using the gossiping policy \cite{yates21gossip} for a comparison. 

\begin{figure}[t]
\centerline{\includegraphics[scale=0.54]{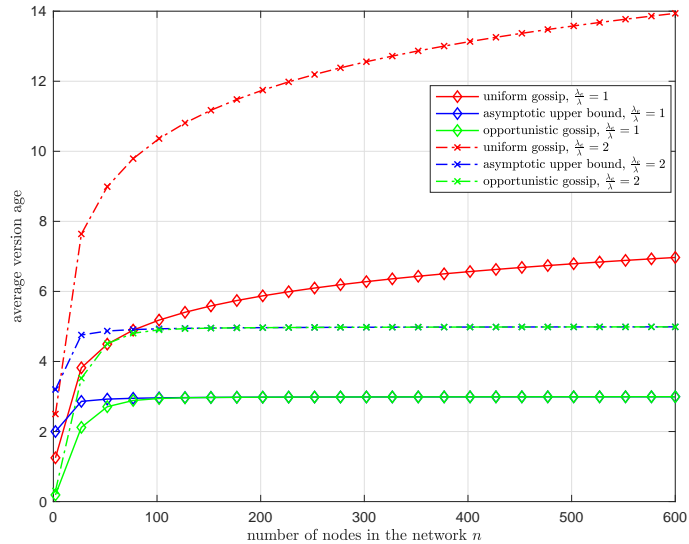}}
\caption{Average version age of a single node versus the total number of nodes in the network $n$.}
\label{plot}
\end{figure}

The results of the simulations are shown in Fig.~\ref{plot}. We observe in Fig.~\ref{plot} that the opportunistic gossiping of ASUMAN performs better than uniform rate gossiping. The uniform rate gossip average age scales as $O(\log n)$, whereas the asymptotic upper bound for the average age in opportunistic gossiping scales as $O(1)$ as proven in Theorem~\ref{thm1}. As calculated from \eqref{final_UB}, the upper bound is 3 and 5, for $\frac{\lambda_{e}}{\lambda}=1$ and $\frac{\lambda_{e}}{\lambda}=2$, respectively. The simulations show that the upper bound is loose when $n$ is small, and it gets tighter as $n$ becomes large. This is expected because the overall network is being updated from the source with rate $\lambda$. Therefore, with large $n$, the update rate of each individual node $\frac{\lambda}{n}$ gets smaller. Hence, in the interval $\mathcal{I}_k$, only a few nodes get updated directly from the source. However, for small $n$, the number of such nodes will be higher. This results in the average node age to be lower than the upper bound in \eqref{final_UB}. Also, we notice that the asymptotic upper bound is an increasing function of $\frac{\lambda_{e}}{\lambda}$. This result matches intuition. If $\frac{\lambda_{e}}{\lambda}$ increases, that means that the source is updating itself more frequently as compared to updating the network. This would result in higher average age. The opposite effect happens when $\lambda$ increases instead of $\lambda_{e}$, thus, resulting in lower average age.

\begin{figure}[t]
\centerline{\includegraphics[scale=0.54]{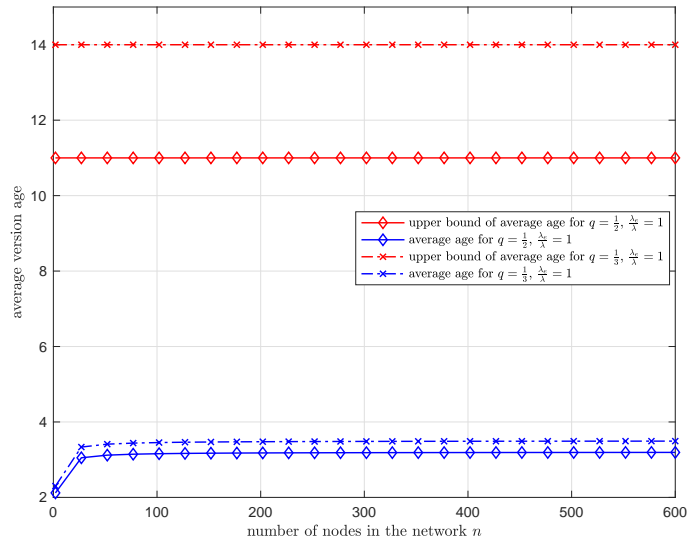}}
\caption{Average version age of a single node versus the total number of nodes $n$ in a partially connected network.}
\label{partial_plot}
\end{figure}

Next, we show the results for partially connected gossip networks. In Fig.~\ref{partial_plot}, we show the average age and theoretical upper bound for $q=\frac{1}{2}$ and $q=\frac{1}{3}$. For both cases, $\frac{\lambda_e}{\lambda}=1$. While the bounds are not tight, from the figure it is clear that the average age of a node in such a network scales as $O(1)$. Also, with decrease in $q$, the average age and the upper bound increases. This is expected, as $q$ becoming smaller implies less connectivity for the network at any fixed time slot, thus increasing the average age of a node.

\begin{figure}[t]
\centerline{\includegraphics[scale=0.54]{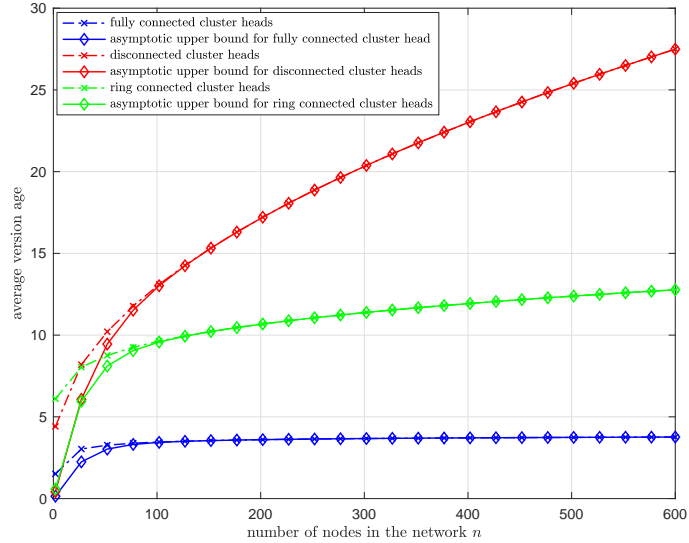}}
\caption{Average version age of a single node versus the total number of nodes for clustered networks with $c=\sqrt{n}$, $p=0.5$.}
\label{cluster_plot}
\end{figure}

Next, we show the simulation results for clustered networks in Fig.~\ref{cluster_plot}. We choose $c=\sqrt{n}$ and $p=\frac{1}{2}$. From the figure, it is evident that the average ages scale as $O(1)$, $O(\sqrt{n})$, and $O(n^{1/4})$, for fully-connected, disconnected and ring-connected cluster head structures, respectively. The disconnected clusters have the highest asymptotic average age scaling due to lack of gossiping among the cluster heads, and the fully-connected clusters have the lowest due to ASUMAN gossiping in both layers. From the graphs, it is clear that the best performance is achieved if all the clusters are merged into a single fully-connected gossip network, and clustering should be as limited as possible for better age performance.

\begin{figure}[t]
\centerline{\includegraphics[scale=0.54]{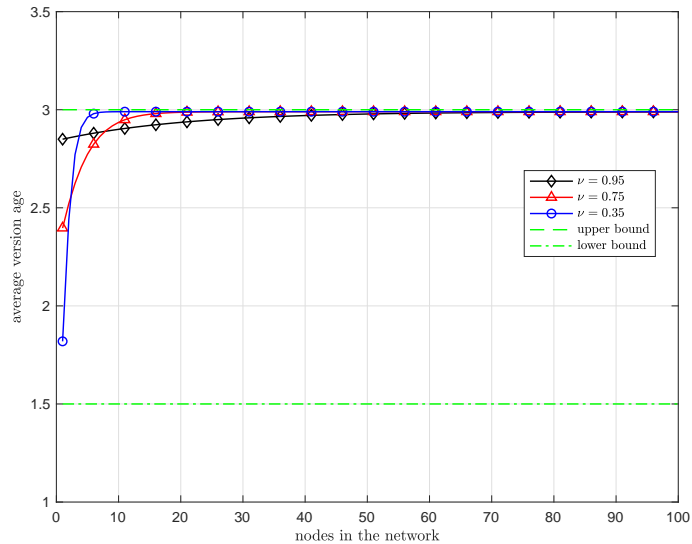}}
\caption{Average version age of nodes in a $n=100$ node fully-connected network with asymmetric power law arrivals.}
\label{power_law_plot}
\end{figure}

Finally, we present an example of asymmetric arrivals in the form of power law update rates. In this simulation, we consider three cases: $\nu=0.95$ and $\nu=0.75$ and $\nu=0.35$ for a network of $100$ nodes with $\frac{\lambda_e}{\lambda}=1$. From Fig.~\ref{power_law_plot}, we observe that all the ages of the nodes are bounded by the upper bound $3$ and lower bound $\frac{3}{2}$, as obtained from Theorem~\ref{thm_asymm}. It can also be observed that the nodes which receive updates less frequently from the source have higher average age as compared to the nodes with more frequent updates. This result follows from the discussion of the age upper bound for power law arrivals in Section~\ref{sect:asymmetric}. For nodes that are close to the $100$th node, the age takes the highest possible value of the upper bound $2\frac{\lambda_e}{\lambda}+1=3$.

\section{Conclusion and Discussion}

We proposed ASUMAN, a gossiping policy for a network of nodes, where the nodes gossip opportunistically instead of uniformly. The network gets synchronized when the source updates itself, and the fresher nodes of the network enter into gossiping phase, following an age sensing phase. This policy allows nodes with relatively fresher versions to gossip with higher rates, and nodes with staler versions to remain silent. We showed that in dense networks, the average age of a node for such a system scales as $O(1)$, which is an improvement compared to gossiping with uniform rates, where the average version age of a node scales as $O(\log n)$.

We further extended our original settings to a system model with partial connections, i.e., only a fixed fraction of edges of a fully-connected network is used for communication at any time slot. We showed that for such a system with ASUMAN gossiping scheme, the average age of a single node still scales as $O(1)$.

Next, we focused on networks with finite connectivity, such as the ring network, two-dimensional grid network, etc., and showed that using ASUMAN for these networks yields $\Omega(n)$ age performance, which is worse than the uniform gossip age performance, pointing to the need for sufficient connectivity for opportunistic gossiping to perform well. 

Then, we modified the network structure and used a hierarchical clustered network, where the connectivity is $O(\sqrt{n})$, which is a trade-off between the fully-connected and finite connected setting and showed that this network has $O(1)$ age. We observed that the hierarchical structure of the network allows some nodes to track their leaders and gossip with ASUMAN locally. This ensures no degradation in the order of age scaling while reducing the order of connectivity.

Next, we considered clustered network models with ring connected and fully-connected cluster heads. We showed that the upper bound to the average age of a cluster node grows linearly with the number of cluster heads $c$ and $\sqrt{c}$ for the two models, respectively. Choosing $c=\sqrt{n}$, the average age scales as $O(\sqrt{n})$ and $O(n^{1/4})$, respectively.

Finally, we considered asymmetric arrivals. We showed that even if the source updates some nodes more frequently than others, due to the ASUMAN scheme, all the nodes of the network have their age scaled as $O(1)$. In particular, we discussed the special case of power law update arrivals and showed that the average age scaling of every node is bounded between two constants.

As a future direction, one may study distributed gossiping schemes that give better performance than uniform gossiping in topologies other than fully-connected networks, such as ring networks, grid networks, etc., and generalize the notion of decentralized gossiping. Another interesting application is the case of multi-source distributed gossiping and its application in various optimization algorithms, and their performance under mobility constraints.

\bibliographystyle{unsrt}
\bibliography{reference}

\end{document}